\documentclass{article}
\usepackage[english]{babel}
\usepackage{amsmath,amssymb,amsfonts,textcomp}
\usepackage[round]{natbib}
\usepackage{bm}
\usepackage{algorithm}
\usepackage{algpseudocode}
\usepackage{multirow}
\usepackage{graphicx}
\usepackage{subfigure}
\usepackage{dsfont}
\usepackage{url}
\usepackage[]{appendix} 

\begin{document}

\title{The Conditional Censored Graphical Lasso Estimator}

\newtheorem{thm}{Theorem}
\newtheorem{dfn}{Definition}
\newtheorem{prop}{Proposition}

\newcommand{\0}{\phantom{0}}
\newcommand{\bmx}{\bm{\mathrm{x}}}
\newcommand{\bmX}{\bm{\mathrm{X}}}
\newcommand{\bmY}{\bm{\mathrm{Y}}}
\newcommand{\bmR}{\bm{\mathrm{R}}}
\newcommand{\bmV}{\bm{\mathrm{V}}}
\newcommand{\mS}{\mathcal S}
\newcommand{\mG}{\mathcal G}
\newcommand{\mV}{\mathcal V}
\newcommand{\mE}{\mathcal E}
\newcommand{\tr}[1]{\mathrm{tr}(#1)}
\newcommand{\diag}[1]{\mathrm{diag}(#1)}
\newcommand{\sign}[1]{\mathrm{sign}(#1)}
\newcommand{\middlemid}{\;\middle\vert\;}

\author{Luigi Augugliaro$^\ast$\\[4pt]
	\textit{Department of Economics, Business and Statistics, University of Palermo, Palermo, Italy}\\[2pt]
	{luigi.augugliaro@unipa.it}\\[4pt]
	Gianluca Sottile\\[4pt]
	\textit{Department of Economics, Business and Statistics, University of Palermo,Palermo, Italy}\\[4pt]
	Veronica Vinciotti\\[4pt]
	\textit{Department of Mathematics, Brunel University London, Uxbridge UB8 3PH, United Kingdom}}	

\maketitle

\begin{abstract}In many applied fields, such as genomics, different types of data are collected on the same system, and it is not uncommon that some of these datasets are subject to censoring as a result of the measurement technologies used, such as data generated by polymerase chain reactions and flow cytometer. When the overall objective is that of network inference, at possibly different levels of a system, information coming from different sources and/or different steps of the analysis can be integrated into one model with the use of conditional graphical models. In this paper, we develop a doubly penalized inferential procedure for a conditional Gaussian graphical model when data can be subject to censoring. The computational challenges of handling censored data in high dimensionality are met with the development of an efficient Expectation-Maximization algorithm, based on approximate calculations of the moments of truncated Gaussian distributions and on a suitably derived two-step procedure alternating graphical lasso with a novel block-coordinate multivariate lasso approach. We evaluate the performance of this approach on an extensive simulation study and on gene expression data generated by RT-qPCR technologies, where we are able to integrate network inference, differential expression detection and data normalization into one model.

keywords: Censored data, censored graphical lasso, conditional Gaussian graphical models, high-dimensional setting, sparsity
\end{abstract}

\section{Introduction~\label{sec:cond_GGM}}

Conditional graphical models, also called conditional random fields, were originally introduced in~\cite{LaffertyEtAl_ICML_01}. Formally, let $\bm y = (y_1,\ldots, y_p)^\top$ and $\bm x = (x_1,\ldots, x_q)^\top$ be $p$- and $q$-dimensional random vectors, respectively, and let $\mG = (\mV, \mE)$ be a graph with vertex set $ \mV = \{1,\ldots,p\}$, indexing only the entries in $\bm y$, and edge set $\mE \subseteq \mV\times\mV$, where $(h,k)\in\mE$ iff there is a directed edge from the vertex $h$ to $k$ in $\mG$. An edge is called undirected if both $(h,k)$ and $(k, h)$ are in $\mE$, and the graph $\mG$ is called undirected if it has only undirected edges, which is the case that we will consider in this paper. Let $\bm z = (\bm x^\top, \bm y^\top)^\top$ and denote by $f(\bm y\mid \bm x;\bm\xi)$ the conditional density function of $\bm y$ given $\bm x$. We shall say that the set $\{\bm z, f(\bm y\mid \bm x;\bm\xi), \mG\}$ is a conditional graphical model when $f(\bm y\mid \bm x;\bm\xi)$ obeys the Markov properties with respect to the graph $\mG$. In the case of an undirected graph, two given random variables in $\bm y$, say $y_h$ and $y_k$, are conditionally independent given $\bm x$ and all the remaining random variables in $\bm y$ iff the unordered pair $(h,k)$ does not belong to the edge set $\mE$.

The conditional Gaussian graphical model, 
is a member of the conditional undirected graphical models based on the assumption that $\bm y\mid\bm x\sim N(\bm\mu_{y\mid x}, \Sigma_{y\mid x})$, where $E(\bm y\mid \bm x) =\bm\mu_{y\mid x}$ and $V(\bm y\mid \bm x) = \Sigma_{y\mid x}$. The inverse of the covariance matrix $\Sigma_{y\mid x}$, denoted by $\Theta_{y\mid x}$, is called precision matrix and its entries have a one-to-one correspondence with partial correlation coefficients. Like in the unconditional Gaussian graphical model, $\Theta_{y\mid x}$  is the parametric tool relating the topological structure of the undirected graph $\mG$ to the factorization of the conditional density of $\bm y$ given $\bm x$. That is, it is possible to show that the random variables $y_h$ and $y_k$ are conditionally independent  given $\bm x$ and  all the remaining variables in $\bm y$ iff the corresponding partial correlation coefficient is zero (see for example~\cite{Lauritzen_book_96}). Consequently, the problem of estimating the edge set of the undirected graph $\mG$ is equivalent to the problem of finding non-zero entries in $\Theta_{y\mid x}$. This property, together with recent advances in statistical methods for sparse estimation, has made these methods particularly appealing for a range of applications, such as genomic and bioinformatics, where different sources of data are typically collected on the same biological system and integrative approaches are on high demand. As a motivating  example, one can consider the combined analysis of genetic and genomic data by \cite{YinEtAl_AOAS_11}. In this kind of experiments both gene expression data, modeled by $\bm y$, and genetic variants, modeled by $\bm x$, are measured on the same subjects. Although a direct analysis of the gene expression data alone can provide some insights into the underlying gene regulation network, such an analysis would ignore the effect of genetic variants on the gene expression, that is, the fact that high correlation among genes could be explained by the effects of shared genetic variants. In this cases, it is more informative to study the graph structure of $\bm y$ conditioned on $\bm x$.

In order to overcome the inferential problems in a high-dimensional setting, that is when $\min\{p, q\}$ is larger than the sample size $n$, penalized methods have recently been proposed in the literature. In particular, motivated by the biological data analysis mentioned above, \cite{YinEtAl_AOAS_11} proposed sparse inference for the conditional Gaussian graphical model defined by:
\begin{equation}\label{eqn:model_par_1}
E(\bm y\mid \bm x) = \bm \beta_0 + \bm\beta^\top\bm x,\quad
V(\bm y\mid\bm x) = \Sigma,
\end{equation}
that is, assuming that $\bm x$ influences the conditional expected value through the $(q + 1)\times p$ regression coefficient matrix $\bm B = (\bm\beta_0, \bm\beta^\top)^\top$ whereas the conditional covariance matrix does not depend on $\bm x$. Denoting with $\Theta = (\theta_{hk})$ the corresponding precision matrix, the density function $\phi(\bm y\mid\bm x;\bm B, \Theta)$ can be written as
\begin{equation}\label{eqn:dens_condGGM}
(2\pi)^{-\frac{p}{2}}|\Theta|^{\frac{1}{2}}\exp\left\{-\frac{1}{2} (\bm y - \bm B^\top\bmx)^\top\Theta(\bm y - \bm B^\top\bmx) \right\},
\end{equation}
where, with a little abuse of notation, we let $\bmx = (1, \bm x^\top)^\top$. Given a sample of independent observations, 
the authors propose the following extension of the well-known graphical lasso (glasso) estimator~\citep{YuanEtAl_BioK_07}:
\begin{equation}\label{eqn:cond_glasso}
\{\bm{\widehat B}, \widehat\Theta\} = \arg\max \frac{1}{n}\sum_{i= 1}^n\log\phi(\bm y_i\mid\bm x_i;\bm B, \Theta) - \lambda \|\bm\beta\|_1 - \rho\|\Theta\|_1^-,
\end{equation}
where $ \|\bm\beta\|_1 = \sum_{h,k}|\beta_{hk}|$ and $\|\Theta\|_1^- = \sum_{h\ne k}|\theta_{hk}|$. In this model, the two tuning parameters $\rho$ and $\lambda$ are used to control, respectively, the amount of sparsity in $\bm{\widehat B}$ and $\widehat\Theta$. As elucidated in~\cite{YinEtAl_AOAS_11}, the first penalty function accounts for the biological expectation that each gene has only a few genetic regulators, i.e., each column of $\bm B$ has only a few non-zero regression coefficients, whereas the second penalty function accounts for the expectation that the precision matrix $\Theta$ is  sparse for genetic networks. It is worth noting that the estimator~(\ref{eqn:cond_glasso}) was independently proposed also in~\cite{RothmanEtAl_JCGS_10}. However, in their work the authors focus only on how to efficiently estimate the regression coefficient matrix when incorporating the covariance information. \cite{Wang_StatSinica_15}, on the other hand, proposes a new algorithm for fitting the estimator~(\ref{eqn:cond_glasso}). This is based on the idea of decomposing the initial problem into a series of simpler conditional problems involving, at each step, the conditional log-likelihood function of each $y_h$ given all the remaining variables. The author also studies the asymptotic properties of the estimator for diverging $p$ and $q$ values. Recently, other works have proposed two-steps procedures for fitting a sparse conditional Gaussian graphical model. Among these, \cite{LiEtAl_JASA_12} propose to first use an initial non-sparse estimator for the conditional variance matrix $\Sigma$ using the theory of reproducing kernel Hilbert spaces, and then,  in the second step of the procedure, the resulting estimate is used to formulate a glasso-type problem for the estimation of $\Theta$. In contrast to this, the two-steps procedure proposed in~\cite{YinEtAl_JMA_13} uses an $\ell_1$-penalized multivariate linear regression model with independent errors for estimating the sparse coefficient regression matrix in the first step, whereas, in the second step the precision matrix is estimated using the standard glasso estimator. Asymptotic results are obtained assuming that the error term is sub-Gaussian. A free distribution approach is instead proposed in ~\cite{CaiEtAl_BioK_13}. In this case, $\bm B$ is first estimated through a constrained $\ell_1$ minimization problem not involving the conditional precision matrix, which can be considered as a multivariate extension of the Danzig selector, and then the  precision matrix is estimated using the method proposed in~\cite{CaiEtAl_JASA_11}.  Finally, \cite{ChenEtAl_JASA_16} propose a two steps procedure where the scaled lasso is used in each step to select the amount of sparsity. The authors also propose theoretical results on how to select the optimal values of the two tuning parameters so as to make the entire procedure tuning free.

Another strand of literature on sparse inference for a conditional Gaussian graphical model is theoretically founded on the more stringent assumption that $\bm x$ is also normally distributed with $E(\bm x) = \bm\mu_x$ and $V(\bm x) = \Sigma_{xx}$. Under this assumption, $\bm z\sim N(\bm\mu_z, \Theta_{zz})$ with expected value and precision matrix partitioned as follows:
\begin{equation}\label{eqn:block_par}
\bm\mu_z = %
\begin{pmatrix}
\bm\mu_x\\
\bm\mu_y
\end{pmatrix} %
\quad\text{and}\quad%
\Theta_{zz} = %
\begin{pmatrix}
\Theta_{xx} & \Theta_{xy}\\
\Theta_{yx} & \Theta_{yy}
\end{pmatrix},
\end{equation}
and the conditional density~(\ref{eqn:dens_condGGM}) can be reparameterized using the following identities:
\[
\bm\beta_0 = \bm\mu_y + \Theta_{yy}^{-1}\Theta_{yx}\bm\mu_x,%
\quad\bm\beta = -\Theta_{xy}\Theta_{yy}^{-1}, %
\Theta = \Theta_{yy}.
\]
Denoting with $\phi(\bm y\mid\bm x;\bm\mu_z, \Theta_{xy}, \Theta_{yy})$ the reparameterized conditional density function~(\ref{eqn:dens_condGGM}), \cite{SohnEtAl_PICAIS_12}, \cite{WytockEtAl_pmlr_13} and \cite{YuanEtAl_IEEETIT_14} independently proposed the following penalized estimator of a sparse conditional Gaussian graphical model:
\begin{equation}\label{eqn:cond_glasso2}
\{\widehat\Theta_{xy}, \widehat\Theta_{yy}\} = \arg\max \frac{1}{n}\sum_{i= 1}^n\log\phi(\bm y_i\mid\bm x_i;\bm{\hat\mu}_z, \Theta_{xy}, \Theta_{yy}) - \lambda \|\Theta_{xy}\|_1 - \rho\|\Theta_{yy}\|_1^-,
\end{equation}
where $\bm{\hat\mu}_z = \sum_{i = 1}^n \bm z_i / n$. Although estimator~(\ref{eqn:cond_glasso2}) is merely based on an alternative parameterization of the conditional density, it enjoys two important advantages over the estimator~(\ref{eqn:cond_glasso}). Firstly, the partial correlation coefficients associated to $\Theta_{xy}$ are the natural parametric tools to infer the conditional independence between $x_h$ and $y_k$ given all the remaining variables, whereas a zero regression coefficient, say $\beta_{hk} = 0$, only implies that $x_h$ has no effect on the expected value of $y_k$, when the level of the remaining predictors is kept fixed. Moreover, the stronger the correlation in $\Theta_{yy}$ is, the more difficult it is to recover the non-zero entries in $\Theta_{xy}$ from $\bm\beta$, which is why \cite{ChiquetEtAl_StatComp_17} suggest to use the first penalty function to sparsify $\Theta_{xy}$ instead of $\bm\beta$. Secondly, the log-likelihood function used in definition~(\ref{eqn:cond_glasso2}) is jointly convex in $\{\Theta_{xy}, \Theta_{yy}\}$ (\citealp{YuanEtAl_IEEETIT_14}) and this facilitates both the optimization and theoretical analysis. Although such advantages can only be viewed in light of the strong distributional assumptions on $\bm z$,  the estimator~(\ref{eqn:cond_glasso2}) has gained significant popularity. For example, \cite{ZhangEtAl_PLOS_14} uses this estimator  to infer genetic networks under SNP perturbations. More recently, \cite{HuangEtAl_IEEE_TKDE_18} propose an extension to model dynamic networks influenced by conditioning input variables whereas~\cite{HuangEtAl_IEEE_TNNLS_18} propose an extension to infer multiple conditional Gaussian graphical models with similar structures.

Despite a widespread literature on sparse conditional Gaussian graphical models, there is a large number of fields in applied research where modern measurement technologies make the use of this graphical model theoretically unfounded, even when the assumption of a multivariate Gaussian distribution is satisfied. An example of this is given by Reverse Transcription quantitative Polymerase Chain Reaction (RT-qPCR), a popular technology for gene expression profiling (\citealp{DerveauxEtAl_Methods_10}). This technique is used to measure the expression of a set of target genes in a given sample through repeated cycles of sequence-specific DNA amplification followed by expression measurements. The cycle at which the observed expression first exceeds a fixed threshold is commonly called the cycle-threshold (\citealp{McCallEtAl_BioInfo_14}). If a target is not expressed, the threshold is not reached after the maximum number of cycles (limit of detection) and the corresponding cycle-threshold is undetermined. For this reason, the resulting data are naturally right-censored and specific methods have been developed to cope with this (\citealp{PipelersEtAl_Plos1_17, McCallEtAl_BioInfo_14}) including graphical modelling approaches for network inference (\citealp{AugugliaroEtAl_BioStat_18}). In order to take into account the censoring mechanism as well as the availability of different sources of information on the same system, in this paper we propose a novel double penalized estimator to infer a sparse conditional Gaussian graphical model with censored data. 
The contributions of this paper are three-fold. Firstly, we introduce a suitable generalization of the  notion of conditional Gaussian graphical model for data subject to censoring, secondly we propose a doubly penalized estimator that accounts for different levels of sparsity of the network of response variables and of that linking the predictors and the responses, and, finally, we develop an efficient Expectation-Maximization algorithm for inference, based, on the one hand, on approximate calculations of the
moments of truncated Gaussian distributions and, on the other hand, on a suitably derived two-step procedure alternating graphical
lasso with a novel block-coordinate multivariate
lasso approach.

The outline of this paper is as follows. In Section~\ref{sec:ccGGM} we introduce the  notion of conditional censored Gaussian graphical model and in Section~\ref{sec:sparse_ccGGM} we present our estimator. In this section we also consider the computational aspects of the proposed approach. The problem of tuning parameter selection is treated in Section~\ref{sec:tuning_selection}, whereas in Section~\ref{sec:sim_studies} we use an extensive simulation study to compare the behaviour of the proposed estimator with existing competitors. Finally, in Section~\ref{sec:realdata}, we study a real gene expression data generated by RT-qPCR technologies,
where we are able to integrate network inference, differential expression and data normalization into one model, and in Section~\ref{sec:conclusions} we draw some conclusions. All proofs are reported in the Appendix.

\section{The conditional censored Gaussian graphical models\label{sec:ccGGM}}

In this section we extend the notion of conditional Gaussian graphical models to the case in which the random variables $\bm y$ and $\bm x$ are subject to specific censoring mechanisms. To this end, let $\bm z = (\bm x^\top, \bm y^\top)^\top$ be a (partially) latent random vector with joint density function $g(\bm z; \bm\xi)$  and let $\bm l = (l_1,\ldots,l_q, l_{q+1},\ldots,l_{q + p})^\top$ and $\bm u = (u_1,\ldots,u_q, u_{q + 1},\ldots,u_{q + p})^\top$, with $l_j < u_j$ for $j=1,\ldots,(q + p)$, be the left and right censoring values, respectively. Then, $z_j$ is observed only if it is inside the closed interval $[l_j, u_j]$, otherwise it is censored from below or above depending on whether $z_j < l_j$ or $z_j > u_j$, respectively. Following~\cite{LittleEtAl_Book_02}, the complete specification of the joint distribution of the  observed data can be obtained using a discrete random vector, denoted by  $R(\bm z;\bm l, \bm u)$, with support $\{-1,0,1\}^{q + p}$ and encoding the censoring patterns. Formally, the $j$th element of $R(\bm z;\bm l, \bm u)$ is defined as $R(z_j; l_j, u_j) = I(z_j>u_j) - I(z_j < l_j)$, where $I(\cdot)$ is the usual indicator function. To simplify our notation, the realization of the random vector $R(\bm z;\bm l, \bm u)$ is denoted by $\bm r$. In this way the index set $\mathcal I = \{1,\ldots,q + p\}$ can be partitioned into $o=\{j\in\mathcal I:r_j = 0\}$, $c^+=\{j\in\mathcal I:r_j = +1\}$ and $c^-=\{j\in\mathcal I:r_j = -1\}$. Furthermore, we shall use the convention that a vector indexed by a set of indices denotes the corresponding subvector, so, for example, the subvector of observed elements in $\bm z$ is denoted by $\bm z_o = (z_j)_{j\in o}$ and, consequently, the observed data is $(\bm z_o^\top, \bm r^\top)^\top$. With such a notation, the joint density function of the observed data is given by
\begin{equation}\label{eqn:obs_pdf}
\varphi(\bm z_o, \bm r;\bm\xi) = \int_{D_c} g(\bm z_o, \bm z_c; \bm\xi)\, \text{d}\bm z_{c}\,  I(\bm l_o \le \bm z_o \le \bm u_o),
\end{equation}
where $c = c^-\cup c^+$ and $D_c = (-\infty, \bm l_{c^-})\times(\bm u_{c^+},+\infty)$.

Similar to the approach described in the Introduction, the notion of conditional censored Gaussian graphical models, formalized below, arises naturally by combining the density~(\ref{eqn:obs_pdf}) with the assumption that the conditional distribution $\bm y$ given $\bm x$ is a multivariate Gaussian distribution.
\begin{dfn}\label{dfn:cond_cGGM}
Let $\bm  z = (\bm x^\top, \bm y^\top)^\top$ be a random vector whose density function can be factorized as follows:
\[
g(\bm z; \bm\xi) = \phi(\bm y\mid\bm x ; \bm\mu_{y\mid x}, \Theta_{y\mid x})\, h(\bm x; \bm\delta),
\]
where $\bm\xi = \{\bm \mu_{y\mid x}, \Theta_{y\mid x},\bm\delta\}$ denotes the full set of parameters, $h(\bm x; \bm \delta)$ denotes the marginal distribution of $\bm x$ and $\phi(\bm y\mid\bm x; \bm\mu_{y\mid x}, \Theta_{y\mid x})$ the conditional distribution of $\bm y$ given $\bm x$, which is assumed to  follow a $p$-dimensional Gaussian distribution. Furthermore, let $R(\bm z;\bm l, \bm u)$ be a $(q + p)$-dimensional random censoring-data indicator defined by the censoring vectors $\bm l$ and $\bm u$. The conditional censored Gaussian graphical model is defined by the set $\{\bm z, R(\bm z; \bm l, \bm u), \varphi(\bm z_o, \bm r;\bm\xi), \mG\}$, where
\begin{equation}\label{eqn:dens_cond_cGGM}
\varphi(\bm z_o, \bm r;\bm\xi) = \int_{D_c} \phi(\bm y_o, \bm y_c\mid \bm x_o, \bm x_c;\bm\mu_{y\mid x}, \Theta_{y\mid x})  \times h(\bm x_o, \bm x_c;\bm\delta) \text{d}\bm x_{c} \text{d}\bm y_{c} 
\end{equation}
with $\bm z_o \in[\bm l_o, \bm u_o]$ and $\phi(\bm y\mid \bm x;\bm\mu_{y\mid x}, \Theta_{y\mid x})$ factorizes according to the undirected graph $\mG = \{\mV,\mE\}$.
\end{dfn}
Although the proposed definition is general enough to cover a broad class of marginal distributions for the vector $\bm x$, in the next section we shall focus on two specific cases. The first one, which we consider in Section~\ref{sec:cond_cglasso_type1}, is the simplest and is obtained when only the random variable $\bm y$ is subject to a censoring mechanism.  Section~\ref{sec:cond_cglasso_type2}, on the other hand, considers the case in which also the random variable $\bm x$ is censored but, in this case, as typically done in literature on conditional Gaussian graphical models,  we enforce  distributional assumptions and require that $\bm x$ is also Gaussian distributed.

\section{Sparse inference on the conditional censored Gaussian graphical models\label{sec:sparse_ccGGM}}

\subsection{Sparse inference when $\bm x$ is fully observed\label{sec:cond_cglasso_type1}}

As proposed in~\cite{YinEtAl_AOAS_11} and~\cite{RothmanEtAl_JCGS_10}, we begin by assuming that $\bm x$ influences the parameters of the conditional distribution of $\bm y$ given $\bm x$ only through a linear model for the conditional expected values, i.e., model~(\ref{eqn:model_par_1}). Then, under the assumption that $\bm x$ is not subject to a censoring mechanism, density~(\ref{eqn:dens_cond_cGGM}) can be significantly simplified as it can be rewritten as the product of the marginal density of $\bm x$ and the conditional density function of the observed response variable $\bm y_o$ given $\bm x$, i.e.:
\begin{eqnarray}\label{eqn:dens_cond_cGGM1}
\varphi(\bm z_o, \bm r;\bm\xi) &=& h(\bm x;\bm\delta) \int_{D_c} \phi(\bm y_o, \bm y_c\mid \bm x;\bm B, \Theta) \text{d}\bm y_{c} \nonumber\\
&=& h(\bm x;\bm\delta)\, \psi(\bm y_o, \bm r\mid \bm x ;\bm B, \Theta),
\end{eqnarray}
with $\bm y_o\in [\bm l_o, \bm u_o ]$. The factorization~(\ref{eqn:dens_cond_cGGM1}) implies that inference about $\{\bm B, \Theta\}$ and $\bm\delta$ can be done separately.  Therefore, since all the relevant information needed to infer  the conditional independence structure of $\bm y$ adjusted for $\bm x$ is contained in $\psi(\bm y_o, \bm r\mid \bm x ;\bm B, \Theta)$, in the following of this section we shall focus only on how to estimate the matrices $\bm B$ and $\Theta$. Inference about $\bm\delta$ can be achieved using, for example, the maximum likelihood method if the sample size is large enough or a suitable  penalized method otherwise.

Suppose that we have a set of $n$ independent observations denoted by $(\bm y_{io_i}^\top, \bm r_i^\top, \bm x_i^\top)^\top$, with $i = 1,\ldots,n$ and $o_i = \{j\in\mathcal I:r_{ij} = 0\}$. Then the relevant average observed log-likelihood function is given by
\begin{equation}\label{eqn:avell}
\bar\ell(\bm B, \Theta) = \frac{1}{n}\sum_{i=1}^n\log \int_{D_{c_i}} \phi(\bm y_{io_i}, \bm y_{ic_i}\mid \bm x_i;\bm B, \Theta) \text{d}\bm y_{ic_i},
\end{equation}
where $c_i = c^-_i\cup c^+_i$, with $c^-_i=\{j\in\mathcal I:r_{ij} = -1\}$ and $c^+_i=\{j\in\mathcal I:r_{ij} = +1\}$. Under a high-dimensional setting, that is $\min\{p, q\} > n$, inference about $\bm B$ and $\Theta$ can be carried out only under the assumption that these matrices have a sparse structure, that is only a few regression coefficients and partial correlation coefficients are different from zero. Then, similarly to the conditional glasso estimator (\ref{eqn:cond_glasso}), we propose to estimate the parameters of a censored conditional Gaussian graphical model by maximizing a new objective function whereby two specific lasso-type penalty functions are added to the average observed log-likelihood. The resulting estimator, which we call conditional cglasso, is defined as follows:
\begin{equation}\label{dfn:cond-cglasso}
\{\bm{\widehat{B}}, \widehat\Theta\} = \arg\max \bar\ell(\bm B, \Theta) - \lambda\sum_{k = 1}^p \theta_{kk}\|\bm\beta_k\|_1- \rho \|\Theta\|_1^{-}
\end{equation}
where $\bm\beta_k$ denotes the $k$th column of $\bm \beta$. Like in the standard conditional graphical lasso estimator, the tuning parameter $\lambda$  is used to control the amount of sparsity in the estimated regression coefficient matrix whereas $\rho$ is devoted to control the sparsity in $\widehat\Theta = (\hat\theta_{hk})$ and, consequently, in the corresponding estimated graph $\widehat \mG = \{\mV, \widehat{\mE}\}$, where $\widehat\mE = \{(h,k):\hat\theta_{hk} \ne0\}$. When $\rho$ is large enough, some  $\hat\theta_{hk}$ are shrunken to zero resulting in the removal of the corresponding link in $\widehat\mG$; on the other hand, when $\rho$ is equal to zero and the sample size is large enough the estimator $\widehat\Theta$ coincides with the maximum likelihood estimator of the precision matrix, which implies a fully connected estimated graph.

A close look at the definition~(\ref{dfn:cond-cglasso}) reveals that the penalty function associated to $\bm B$ is different to what is proposed in the literature \citep{RothmanEtAl_JCGS_10,YinEtAl_AOAS_11}. In particular, we use the diagonal elements of the concentration matrix to scale the $\ell_1$-norm of the parameter vectors $\bm\beta_k$. As we shall see in the next section, the use of these scaling factors increases significantly the computational efficiency of the proposed algorithm.

\subsection{An efficient Expectation-Maximization algorithm\label{sec:em-algo1}}

In order to address the computational burden in the computation of the integral involved in the definition of the average log-likelihood function~(\ref{eqn:avell}) and to  solve efficiently the maximization problem in definition~(\ref{dfn:cond-cglasso}), we develop an efficient Expectation-Maximization (EM) algorithm \citep{DempsterEtAl_JRSSB_77}. The EM algorithm is based on the idea of repeating two steps until a convergence criterion is met.

The first step, called E-Step, requires the calculation of the conditional expected value of the complete log-likelihood function using the current estimates. As shown in \cite{McLachlanEtAl_Book_08}, this step can be significantly simplified when the complete probability density function is a member of the regular exponential family since, in this case, the E-Step simply requires the computation of the conditional expected values of the sufficient statistics. In our model, the E-Step is theoretically founded on the following theorem, which relates the partial derivatives of the average observed log-likelihood function~(\ref{eqn:avell}) to the expected value operator computed with respect to the conditional distribution of $\bm y_{ic_i}$ given $\{\bm x_i, \bm y_{io_i}\}$ and truncated over a specific region. For the sake of simplicity,  in the remaining part of this paper, when needed,  we let $\bm\vartheta = \{\bm B, \Theta\}$ and we let $\vartheta_{hk}$ denote a generic parameter.

\begin{thm}\label{prop:main_identity}
Let $E(\;\cdot\mid \bm y_{ic_i}\in D_{c_i}, \bm x_i; \bm\vartheta)$ be the expected value operator computed with respect to the conditional Gaussian distribution of $\bm y_{ic_i}$ given $\{\bm x_i, \bm y_{io_i}\}$ and truncated over the region $D_{c_i}$. Then $\partial \bar\ell(\bm\vartheta)/\partial\vartheta_{hk}$ is equal to:
\[
\frac{1}{n}\sum_{i= 1}^n E\left(\frac{\partial\log\phi(\bm y_{io_i}, \bm y_{ic_i} \mid \bm x_i;\bm\vartheta)}{\partial\vartheta_{hk}} \middlemid \bm y_{ic_i}\in D_{c_i},\bm x_i;\bm{\vartheta}\right).
\]
\end{thm}
Let $\bm{\hat\vartheta}$ be the current estimate of $\bm\vartheta$. Let $\widehat\bmY = (\hat y_{i,k})$, with
\[
\begin{array}{rl}
\hat y_{i,k} & = \left\{%
\begin{array}{ll}
y_{ik} & \mbox{ if } r_{ik} = 0\\
E(y_{ik}\mid \bm y_{ic_i}\in D_{c_i}, \bm x_i;\bm{\hat\vartheta}) & \mbox{ otherwise},
\end{array}
\right.%
\end{array}
\]
representing the current $(n\times p)$-dimensional matrix of the imputed response variables, obtained by replacing the right and left censoring values by the expected values of the truncated normal distribution. In the same way, we let $\bm{\widehat C}_{yy} = (\sum_{i=1}^n \hat y_{i,hk})$ be the $(p\times p)$-dimensional matrix where $\hat y_{i,hk}$ is equal to:
\[
\begin{array}{ll}
y_{ih} y_{ik}& \mbox{ if } r_{ih} = 0 \mbox { and } r_{ik} = 0\\
y_{ih} E(y_{ik}\mid \bm y_{ic_i}\in D_{c_i}, \bm x_i; \bm{\hat\vartheta}) & \mbox{ if } r_{ih} = 0 \mbox { and } r_{ik} \ne 0\\
E(y_{ih}\mid \bm y_{ic_i}\in D_{c_i}, \bm x_i; \bm{\hat\vartheta}) y_{ik} & \mbox{ if } r_{ih} \ne 0 \mbox { and } r_{ik} = 0\\
E(y_{ih} y_{ik}\mid \bm y_{ic_i}\in D_{c_i}, \bm x_i; \bm{\hat\vartheta}) & \mbox{ if } r_{ih} \ne 0 \mbox { and } r_{ik} \ne 0.\\
\end{array}
\]
With such notation, the E-Step of the proposed algorithm requires the computation of the imputed empirical conditional covariance matrix:
\[
\bm{\widehat S}_{y\mid x} (\bm B) = n^{-1}\{\bm{\widehat C}_{yy} - \bm{\widehat C}_{yx} \bm B - \bm B^\top\bm{\widehat C}_{xy} + \bm B^\top \bm C_{xx} \bm B\},
\]
where $\bm{\widehat C}_{yx} = \widehat\bmY^\top\bmX$, $\bm{\widehat C}_{xy} = \bm{\widehat C}_{yx}^\top$, $\bm C_{xx} = \bmX^\top\bmX$ and $\bmX = (\bmx_1,\ldots,\bmx_n)^\top$ denotes the usual $n\times (q+1)$ design matrix.

The second step of the EM algorithm, the M-Step, requires the solution of a new maximization problem obtained by replacing the objective function of definition~(\ref{dfn:cond-cglasso}), with the so-called penalized $Q$-function:
\begin{equation}\label{eqn:Qfun_cond_cglasso_type1}
Q(\bm B, \Theta) = \log\det\Theta - \mathrm{tr}\{\Theta\bm{\widehat S}_{y\mid x}(\bm B)\} - \lambda \sum_{k = 1}^p \theta_{kk} \|\bm\beta_k\|_1   - \rho  \|\Theta\|^-_1.
\end{equation}
The problem of maximizing this function is formally equivalent to the problem studied in \cite{RothmanEtAl_JCGS_10} and \cite{YinEtAl_AOAS_11}. In particular, since for a fixed $\bm{\hat\vartheta}$ the penalized $Q$-function~(\ref{eqn:Qfun_cond_cglasso_type1}) is a bi-convex function of $\bm B$ and $\Theta$, its maximization can be obtained by repeating two sub-steps until a convergence criterion is met. In the first sub-step, given $\widehat\Theta$, the regression coefficient matrix is estimated by solving the following sub-problem:
\begin{equation}\label{eqn:cond_cglasso_subprob1}
\min_{\bm B} \mathrm{tr}\{\widehat\Theta\bm{\widehat{S}}_{y\mid x}(\bm B)\} + \lambda \sum_{k = 1}^p \hat\theta_{kk} \|\bm\beta_k\|_1,
\end{equation}
whereas, in the second sub-step, given $\bm{\widehat{B}}$, the matrix $\Theta$ is estimated by solving the sub-problem:
\begin{equation}\label{eqn:cond_cglasso_subprob2}
\max_{\Theta\succ0} \log\det\Theta - \mathrm{tr}\{\Theta \bm{\widehat{S}}_{y\mid x}(\bm{\widehat{B}})\} - \rho  \|\Theta\|^-_1.
\end{equation}
Algorithm~\ref{algo:EM_cond_cglasso_type1} reports the pseudo-code of the proposed EM algorithm.
\begin{algorithm}[t]
\caption{Pseudo-code of the proposed EM algorithm\label{algo:EM_cond_cglasso_type1}}
\begin{algorithmic}[1]
\State Let $\bm{\widehat{B}}$ and $\widehat\Theta$ be the initial estimates
\Repeat
\State let $\bm{\hat\vartheta} = \{\bm{\widehat{B}}, \widehat\Theta\}$
\State update $\widehat\bmY$, $\bm{\widehat{C}}_{yy}$ \Comment{E-Step}
\Repeat\Comment{M-Step}
\State $\bm{\widehat{B}} = \arg\min_{\bm B} \mathrm{tr}\{\widehat\Theta\bm{\widehat{S}}_{y\mid x}(\bm B)\} + \lambda \sum_{k = 1}^p \hat\theta_{kk}\|\bm\beta_k\|_1$ 	
\State $\widehat\Theta = \arg\max_{\Theta\succ0} \log\det\Theta - \mathrm{tr}\{\Theta\bm{\widehat{S}}_{y\mid x}(\bm{\widehat{B}})\} - \rho  \|\Theta\|^-_1$
\Until{convergence criterion is met}
\Until{convergence criterion is met}
\end{algorithmic}
\end{algorithm}
While problem~(\ref{eqn:cond_cglasso_subprob2}) is a standard graphical lasso problem that can be efficiently solved using, for example, the block-coordinate descent algorithm proposed in~\cite{FriedmanEtAl_biostat_08},  problem~(\ref{eqn:cond_cglasso_subprob1}) is similar to what is studied by \cite{RothmanEtAl_JCGS_10} and \cite{YinEtAl_AOAS_11} in the case of no censoring. However, instead of solving this problem through a cyclic coordinate descent algorithm, we propose a more efficient and easy-to-implement block-coordinate descent algorithm. In particular, the algorithm that we propose in this paper is based on the following result.
\begin{thm}\label{prop:multi_lasso}
Let $\bm{\widehat{S}}_{y\mid x}(\bm B_k)$ be the matrix $\bm{\widehat{S}}_{y\mid x}(\bm B)$ seen as function of the $k$-th column of $\bm B$, denoted by $\bm B_k$, while the remaining columns are held fixed to the current estimates. Then, the following minimization problem
\[
\min_{\bm B_k}  \mathrm{tr}\{\widehat\Theta\bm{\widehat{S}}_{y\mid x}(\bm B_k)\} + \lambda  \hat\theta_{kk} \|\bm\beta_k\|_1,
\]
is equivalent to
\[
\min_{\bm B_k} \frac{1}{n} \|\widetilde\bmY_k - \bmX\bm B_k \|^2  + \lambda\|\bm\beta_k\|_1,
\]
where $\widetilde\bmY_k$ is a vector with $i$th element $\tilde y_{i,k} = \hat y_{i,k} +\hat\theta_{kk}^{-1} \sum_{h\neq k}^p\hat\theta_{hk}\{\hat y_{i,h} - \bmx_i^\top\bm{\widehat{B}}_h\} $.
\end{thm}
Theorem~(\ref{prop:multi_lasso}) reveals that the maximization of the penalized $Q$-function~(\ref{eqn:Qfun_cond_cglasso_type1}) with respect to $\bm B_k$, while the remaining parameters are held fixed, is equivalent to a standard lasso problem~\citep{Tibshirani_JRSSB_96} that can be efficiently solved using the algorithm proposed in~\cite{FriedmanEtAl_JSS_10}. Algorithm~\ref{algo:multi_lasso} reports the pseudo-code of the method proposed to solve sub-problem~(\ref{eqn:cond_cglasso_subprob1}), which we call multi-lasso algorithm. Convergence to a global minimizer follows from the fact that the trace term in sub-problem~(\ref{eqn:cond_cglasso_subprob1}) is a convex and differentiable function and the penalty function can be decomposed as sum of $p$ convex functions~\citep{Tseng_JOTA_01}. Furthermore, we underline that the computational efficiency of the proposed multi-lasso algorithm can be improved using the  results given in~\cite{WittenEtAl_JCGS_11}. In other words, when $\rho$ is large enough, and after a permutation of the response variables, $\widehat\Theta$ has a block diagonal structure. Therefore, Theorem~\ref{prop:multi_lasso} implies that $\bm{\widehat{B}}$ can be computed by running  the multi-lasso algorithm in parallel. This insight makes the proposed algorithm useful for the analysis of data sets where the number of predictors is really large.
\begin{algorithm}[t]
\caption{Pseudo-code of the proposed multi-lasso algorithm\label{algo:multi_lasso}}
\begin{algorithmic}[1]
\State Let $\widehat\Theta$ be the current estimate of the precision matrix
\Repeat
\For{$k = 1\ldots p$}
\State compute $$\tilde y_{i,k} = \hat y_{i,k} + \hat\theta_{kk} ^{-1}\sum_{h\neq k}^p\hat\theta_{hk}\{\hat y_{i,h} - \bmx_i^\top\bm{\widehat{B}}_h\} $$
\State compute $$ \bm{\widehat{B}}_k = \arg\min_{\bm B_k} \frac{1}{n} \| \widetilde\bmY_k - \bmX \bm B_k \|^2  + \lambda\|\bm\beta_k\|_1$$
\State replace the $k$th column of the matrix $\bm{\widehat{B}}$ with $\bm{\widehat B_k}$
\EndFor
\Until{convergence criterion is met}
\end{algorithmic}
\end{algorithm}

Finally, as discussed also in~\cite{AugugliaroEtAl_BioStat_18} in the context of an (unconditional) Gaussian graphical model under censoring, the computational burden needed to compute the mixed moments in $\bm{\widehat{C}}_{yy}$ can be too extensive when the estimated precision matrix is dense. For this reason, also in this paper, we propose an approximate EM algorithm defined by replacing the matrix $\bm{\widehat{C}}_{yy}$ with the approximation suggested in~\cite{GuoEtAl_JCGS_15}. By an extensive simulation study, \cite{AugugliaroEtAl_BioStat_18} shows that such an approximation works well no a similar problem and for a large range of $\rho$-values.

\subsection{Extension to the censored $\bm x$ case\label{sec:cond_cglasso_type2}}

Although the estimator proposed in Section~\ref{sec:cond_cglasso_type1} is developed under the assumption that the random vector of predictors is fully observed,  the approach can be easily extended to the case in which both $\bm y$ and $\bm x$ are subject to censoring mechanisms. Following the approaches described in the Introduction, in this section we enforce a distributional assumptions and assume that $\bm x$ is also normally distributed with $E(\bm x) = \bm\mu_x$ and $V(\bm x) = \Sigma_{xx}$. Under this assumption, the likelihood function of the observed data set $(\bm z_o^\top, \bm r^\top)^\top$ is formally equivalent to the likelihood function of the censored Gaussian graphical model proposed in~\cite{AugugliaroEtAl_BioStat_18}, i.e.:
\[
\varphi(\bm z_o, \bm r; \bm\mu_z,\Theta_{zz}) = \int_{D_c} \phi(\bm z_o, \bm z_c; \bm\mu_z,\Theta_{zz})\, \text{d}\bm z_{c}.
\]
Given a set of $n$ independent observations, we therefore propose to estimate the sub-matrices of $\Theta_{zz}$ through the following penalized estimator:
\begin{equation}\label{dfn:cond-cglasso_2}
\{\widehat\Theta_{xy}, \widehat\Theta_{yy}, \widehat\Theta_{xx}\} = \arg\min \frac{1}{n}  \sum_{i = 1}^n\log \varphi(\bm z_{io_i}, \bm r_i; \bm\mu_z,\Theta_{zz}) - \lambda \|\Theta_{xy}\|_1  - \rho(\|\Theta_{xx}\|_1^- + \Theta_{yy}\|_1^-).
\end{equation}
Estimator~(\ref{dfn:cond-cglasso_2}) is an extension of the censored glasso estimator of~\cite{AugugliaroEtAl_BioStat_18}, now specifically designed to estimate separately the submatrices $\Theta_{xy}, \Theta_{yy}$ and $\Theta_{xx}$ in order to emphasize the different role that they have in this kind of graphical model. It is worth noting that, differently to the approaches without censored data, the log-likelihood function used to define the estimator~(\ref{dfn:cond-cglasso_2}) depends on the submatrix $\Theta_{xx}$ and therefore this matrix cannot be ignored  in the fitting process. More specifically, an accurate estimate of $\Theta_{xx}$ is needed in the E-Step to compute the conditional expected values of the sufficient statistics. Finally, although estimator~(\ref{dfn:cond-cglasso_2}) is defined using the same tuning parameter to control the amount of sparsity in $\Theta_{xx}$ and $\Theta_{yy}$, the definition can be generalized by using two specific tuning parameters, say $\rho_1$ and $\rho_2$. However, such a choice would increase significantly the computational burden of selecting the optimal values of these tuning parameters, making the application of the proposed estimator infeasible in the case of a large dataset.

\section{Tuning parameters selection\label{sec:tuning_selection}}

The two tuning parameters play a central role in the proposed conditional cglasso estimator~(\ref{dfn:cond-cglasso}) since they are designed to control the complexity of the topological structure of the estimated graph and the effects of the predictors on the response variables. Standard search strategies proposed in the literature to select the optimal values of the tuning parameters of any penalized estimator are structured in two steps, each of which requires specific choices that must be carefully evaluated in order to reduce the computational burden of the global strategy.

The first step requires the definition of a two dimensional grid, say $\{\lambda_{\min},\ldots,\lambda_{\max}\}\times\{\rho_{\min},\ldots,\rho_{\max}\}$, over which to evaluate the goodness-of-fit of the model. When the sample size is large enough, one can let $\lambda_{\min}$ and/or $\rho_{\min}$ equal to zero, so that the maximum likelihood estimate is one of the points belonging to the coefficient path. On the other hand, when we are in a high-dimensional setting, one can use two values small enough to avoid the overfitting of the model. With respect to the largest values of the two tuning parameters, the next theorem gives both the exact formulas for computing $\lambda_{\max}$ and $\rho_{\max}$ and the corresponding conditional cglasso estimates.
\begin{thm}\label{prop:max_lambdarho}
Let $\phi(y_k;\mu_k,\sigma^2_k)$ be the marginal density function of the random variable $y_k$, where $E(y_k) = \mu_k$ and $V(y_k) = \sigma^2_k$. For any index $k$, define the sets $o_k = \{i : r_{ik} = 0\}$, $c^{-}_k = \{i : r_{ik} = -1\}$ and $c^{+}_k = \{i : r_{ik} = +1\}$ and compute the marginal maximum likelihood estimates $\{\hat\mu_k,\hat\sigma^2_k\}$ by maximizing the following log-likelihood function:
\begin{equation*}
\ell(\mu_k,\sigma^2_k) =  \sum_{i\in o_k} \log\phi(y_{ik};\mu_k,\sigma^2_k) + |c^{-}_k|\log\int_{-\infty}^{l_k} \phi(y;\mu_k,\sigma^2_k) \text{d} y + |c^{+}_k|\log\int_{u_k}^{+\infty} \phi(y;\mu_k,\sigma^2_k) \text{d} y.
\end{equation*}
Then, the estimators $\bm{\widehat B} = (\bm{\hat\beta}_0,\bm{\hat\beta}^\top)^\top$ and $\widehat\Theta$ corresponding to the pair $(\lambda_{\max}, \rho_{\max})$ are $\bm{\hat\beta}_0 = (\hat\mu_1,\ldots,\hat\mu_p)^\top$, $\bm{\hat\beta} = \bm0$ and $\widehat\Theta = \diag{\hat\sigma^{-2}_{1}, \ldots, \hat\sigma^{-2}_{p}}$. %
%
%
Furthermore,
\begin{align*}
\lambda_{\max} & = \max_{h,k} n^{-1} |\sum_{i = 1}^m x_{ih}(\hat y_{i,k} - \hat\mu_k)|, \\
\rho_{\max} & = \|\bm{\widehat S}_{y\mid x}(\bm{\widehat B})\|^-_\infty,
\end{align*}
where $\|\bm{\widehat S}_{y\mid x}(\bm{\widehat B})\|^-_\infty = \max_{h\ne k}|\hat s_{hk}(\bm{\widehat B})|$ and $\hat s_{hk}(\bm{\widehat B})$  denotes the generic entry of $\bm{\widehat S}_{y\mid x}(\bm{\widehat B})$.
\end{thm}
The previous results can be easily extended to the case in which also the predictor vector $\bm x$ is partially observed. In this case, the largest values of the two tuning parameters of the estimator~(\ref{dfn:cond-cglasso_2}) can be obtained through a generalization of the results given in~\cite{AugugliaroEtAl_BioStat_18}. In particular, denoting $\{\hat\mu_k, \hat\sigma^2_k\}$ the values of the maximization of the following marginal log-likelihood function:
%
\begin{equation*}
\ell(\mu_k,\sigma^2_k) =  \sum_{i\in o_k} \log\phi(z_{ik};\mu_k,\sigma^2_k)  + |c^{-}_k|\log\int_{-\infty}^{l_k} \phi(z;\mu_k,\sigma^2_k) \text{d} z + |c^{+}_k|\log\int_{u_k}^{+\infty} \phi(z;\mu_k,\sigma^2_k) \text{d} z,
\end{equation*}
%
and by
\[
\bm{\widehat S}_{zz}(\bm{\hat\mu}_z, \widehat\Theta_{zz}) = %
\begin{pmatrix}
\bm{\widehat S}_{xx}(\bm{\hat\mu}_z, \widehat\Theta_{zz}) & \bm{\widehat S}_{xy}(\bm{\hat\mu}_z, \widehat\Theta_{zz})\\
\bm{\widehat S}_{yx}(\bm{\hat\mu}_z, \widehat\Theta_{zz}) & \bm{\widehat S}_{yy}(\bm{\hat\mu}_z, \widehat\Theta_{zz})
\end{pmatrix},
\]
the corresponding imputed empirical covariance matrix of $\bm z$, then
%
\begin{align*}
\lambda_{\max} & =  \| \bm{\widehat S}_{xy}(\bm{\hat\mu}_z, \widehat\Theta_{zz}) \|_\infty, \\%
\rho_{\max} & = \max\{\| \bm{\widehat S}_{xx}(\bm{\hat\mu}_z, \widehat\Theta_{zz}) \|^-_\infty, \| \bm{\widehat S}_{yy}(\bm{\hat\mu}_z, \widehat\Theta_{zz}) \|^-_\infty \}.
\end{align*}
%
Once $\lambda_{\max}$ and $\rho_{\max}$ are calculated, the entire coefficient path can be computed over a two dimensional grid $\{\lambda_{\min},\ldots,\lambda_{\max}\}\times\{\rho_{\min},\ldots,\rho_{\max}\}$ using the estimate obtained for a given pair $(\lambda, \rho)$ as warm starts for fitting the next conditional cglasso model. This strategy is commonly used also in other efficient lasso algorithms and \texttt{R} packages (\citealp{FriedmanEtAl_JSS_10}, \citealp{glasso_manual}).  Motivated by our simulation study, showing the inter-play between the estimation of $\bm B$ and $\bm \Theta$, we suggest the following strategy to compute the coefficient path: starting with the model fitted at $(\lambda_{\max}, \rho_{\max})$, the other models are fitted reducing $\rho$ faster than $\lambda$, that is, keeping fixed $\lambda$ we reduce $\rho$ from $\rho_{\max}$ to $\rho_{\min}$. When $\rho$ is equal to $\rho_{\min}$, we repeat the previous procedure using the next $\lambda$-value in $\{\lambda_{\min},\ldots,\lambda_{\max}\}$. The coefficient path is completed when we fit the model at $(\lambda_{\min}, \rho_{\min})$.

The second step of a tuning strategy requires the definition of a criterion for evaluating the behaviour of the fitted models. Different choices are suggested in the literature: \cite{YinEtAl_AOAS_11}, \cite{StadlerEtAl_StatComp_12} and \cite{LiEtAl_JASA_12} suggest to use the Bayesian Information Citerion (BIC) whereas  \cite{RothmanEtAl_JCGS_10}, \cite{LeeEtAl_JMA_2012} and \cite{CaiEtAl_BioK_13} suggest to use cross-validation. In this paper, we propose to select the optimal values of the two tuning parameters by using the BIC measure, defined as
\begin{equation*}
\hbox{BIC}(\lambda, \rho) = -2 \sum_{i=1}^{n}\log\int_{D_{c_i}} \phi(\bm y_{io_i}, \bm y_{ic_i}\mid \bm x_i;\widehat{\bm B}, \widehat\Theta) \text{d}\bm y_{ic_i} + k(\lambda, \rho) \log n,
\end{equation*}
where $ k(\lambda, \rho)$ denotes the number of non-zero parameters estimated by the proposed method. Although a search on the two dimensional grid defined in the first step can be performed to select the pair $(\lambda, \rho)$ that minimizes $\hbox{BIC}(\lambda, \rho)$, the computational burden related to the evaluation of the log-likelihood function can make this strategy infeasible also for moderate size problems. For this reason, following \citet{IbrahimEtAl_JASA_08}, we propose to select the tuning parameters by minimizing the following approximate measure:
\begin{equation*}
\overline{\hbox{BIC}}(\lambda, \rho) =  - n[\log\det\widehat\Theta_{\widehat\mE} - \mathrm{tr}\{\widehat\Theta_{\widehat\mE}\bm{\widehat S}_{y\mid x}(\widehat{\bm B}_{\widehat\mS})\}] + k(\rho, \lambda) \log n,
\end{equation*}
i.e., by substituting the exact log-likelihood function with the $Q$-function used in the M-Step of the proposed algorithm, which is easily obtained as a byproduct of the EM algorithm.

\section{Simulation studies\label{sec:sim_studies}}

The global behavior of the proposed estimator~(\ref{dfn:cond-cglasso}), which we call conditional cglasso, is compared with the conditional MissGlasso~(\citealp{StadlerEtAl_StatComp_12}), where $\ell_1$-penalized estimation is performed under the assumption that unobserved data are missing at random, and with the conditional glasso~(\ref{eqn:cond_glasso}), where missing data are imputed using the censoring values. The three different methods will be evaluated on the basis of criteria that measure their performance throughout the entire coefficient path. In this way, on the one hand, we avoid  pitfalls of specific tuning parameter strategies, which may favour one estimator over another, and, on the other hand, if  an estimator is found which is uniformly better than the other estimators across the entire path, that is for any possible pair $(\lambda, \rho)$, then such an estimator will be preferable regardless of the method used to select the tuning parameters.

As in the dataset studied in Section~\ref{sec:realdata}, we consider right-censored response variables with $u_k$ equal to 50, for any $k = 1, \ldots, p$.  To simulate a sample of size $n$ from a sparse conditional censored Gaussian graphical model we use the following procedure. First, $\bm x_i$ is simulated from a multivariate Gaussian distribution with zero expected value and covariance matrix obtained using the method implemented in the \texttt{R} package \texttt{huge} (\citealp{huge_manual}); more specifically, we simulate a random structure where the probability that a pair of nodes has an edge is equal to 0.2.

Then, the right-censored response vector $\bm y_i$ is simulated in analogy with the right-censored multivariate linear regression models.
Firstly, we simulate the $p$-dimensional random vector of errors $\bm\epsilon_i$ from a multivariate Gaussian distribution with zero expected value and concentration matrix $\Theta$. The diagonal entries are fixed to 1 while the non-zero partial correlation coefficients $\theta_{h(h + j)}$, with $h = 1, 6, 11, \ldots, p$ and $j = 1, \ldots,4$, are sampled from a uniform distribution on the interval $[0.30, 0.35]$. In terms of graph theory what  we have is a set of stars, i.e.,  complete bipartite graphs with only one internal vertex and four leaves. A similar setting for simulating the sparse concentration matrix was used in~\cite{LiEtAl_JASA_12}. Then, for each $k \in\mV$, we randomly draw the set $\mS_k\subset\{1,\ldots,q\}$, with $|\mS_k | = 2$; the corresponding regression coefficients $\beta_{hk}$, with $h\in\mS_k$, are sampled from a uniform distribution on the interval $[0.3, 0.7]$. We use the intercepts $\beta_{0k}$ to fix the probability of right-censoring. More specifically, $\beta_{0k}$ is chosen in such a way that $\pi = \sum_{i=1}^n\Pr\{y_{ik}\ge u_k\} / n$ is equal to $0.4$ for the first $K$ response variables whereas it is equal to $10^{-6}$ for the remaining response variables. In our study, $K$ is used to evaluate the effect of the number of censored response variables on the behaviour of the considered estimators. Finally, the censored response vector is obtained by the model $\bm y_i = \bm B^\top\bmx_i + \bm \epsilon_i$ and treating each $y_{ik}$ greater than $u_k$ as a right censored value. We perform a simulation study, where the sample size is fixed to $100$ and the quantities $p, q$ and $K$ are used to define the different scenarios under which evaluate the competing estimators. We let $p, q\in\{50, 200\}$, so that we can  consider both a low-dimensional setting, when $p$ and $q$ are equals to $50$, and different high-dimensional settings. Finally, we let $K =   c\,p$, with $c\in\{0.2, 0.4\}$.

As explained earlier, our aim is the evaluation of the entire coefficient path of each competitor. To this end, in order to study the coefficient path for $\widehat\Theta$, we first control the amount of shrinkage on $\bm B$ by keeping fixed the ratio $\lambda/\lambda_{\max}$ and then, for this fixed $\lambda$ value, the path for $\widehat\Theta$ is computed across a decreasing sequence of ten evenly spaced $\rho$-values, from $\rho_{\max}$ to $\rho_{\min} = 0.1\times\rho_{\max}$. The resulting path is evaluated both in term of network recovery and parameter estimation. For the first aspect, we use the precision-recall curve which shows the relationship between the quantities:
\begin{align*}
\hbox{Precision} &=  \frac{\hbox{number of $\hat\theta_{hk}\ne0$ and $\theta_{hk}\ne0$}}{\hbox{number of  $\hat{\theta}_{hk}\ne0$}}, \\
\hbox{Recall}&= \frac{\hbox{number of $\hat\theta_{hk}\ne0$ and $\theta_{hk} \ne 0$}}{\hbox{number of $\theta_{hk}\ne0$}}.
\end{align*}
An estimator is defined globally preferable in terms of sparse recovery when its precision-recall curve is always over the other curves. To simplify our comparison, we summarize the precision-recall curve by using the area under the curve~(AUC). The second aspect is evaluated in terms of mean squared error, defined as
\[
\hbox{MSE}(\widehat\Theta) = E(\|\widehat\Theta - \Theta\|^2_F),
\]
where $\|\cdot\|^2_F$ denotes the squared Frobenius norm. An estimator is defined globally preferable in terms of mean square error when its $\hbox{MSE}(\widehat\Theta)$ path, obtained across $\rho$ values, is below that of the other paths. In order to summarize a given $\hbox{MSE}(\widehat\Theta)$ path, we use the minimum mean squared error attained along the path, which we denote by $\min_\rho\hbox{MSE}(\widehat\Theta)$. Finally, in order to study the effects of the amount of shrinkage applied on $\bm B$ on the different measures, we repeat the procedure across four possible values for the ratio $\lambda/\lambda_{\max}$, i.e., $\lambda/\lambda_{\max}\in\{1.00, 0.75, 0.50, 0.25\}$, whereby the first value corresponds to the case where the predictors are not included in the model. The comparison of the $\bm{\widehat{B}}$ paths for the different estimators is done using the same strategy but inverting the role of $\widehat\Theta$ with $\bm{\widehat{B}}$ and $\rho$ with $\lambda$. The tables reporting the full results of our study, based on 50 replicates, can be found in the Supplementary Materials. Below, we summarize the results in a number of ways.

\begin{figure}
\centering
\subfigure[Precision-Recall Curves for $\bm{\widehat{B}}$ with $\rho/\rho_{\max} = 0.25$\label{fig:pr-curve-B}]
{\includegraphics[scale = 0.20]{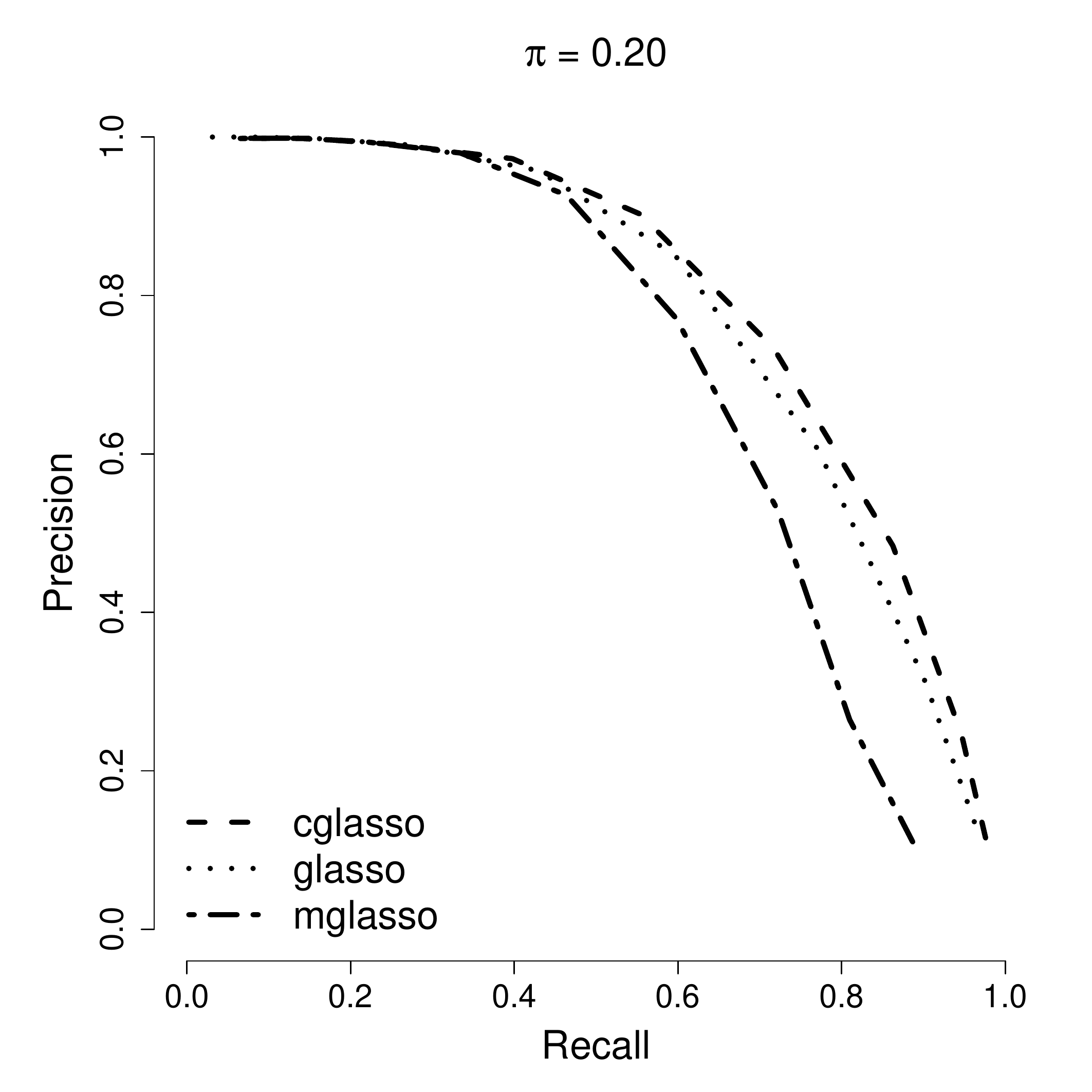}
\includegraphics[scale = 0.20]{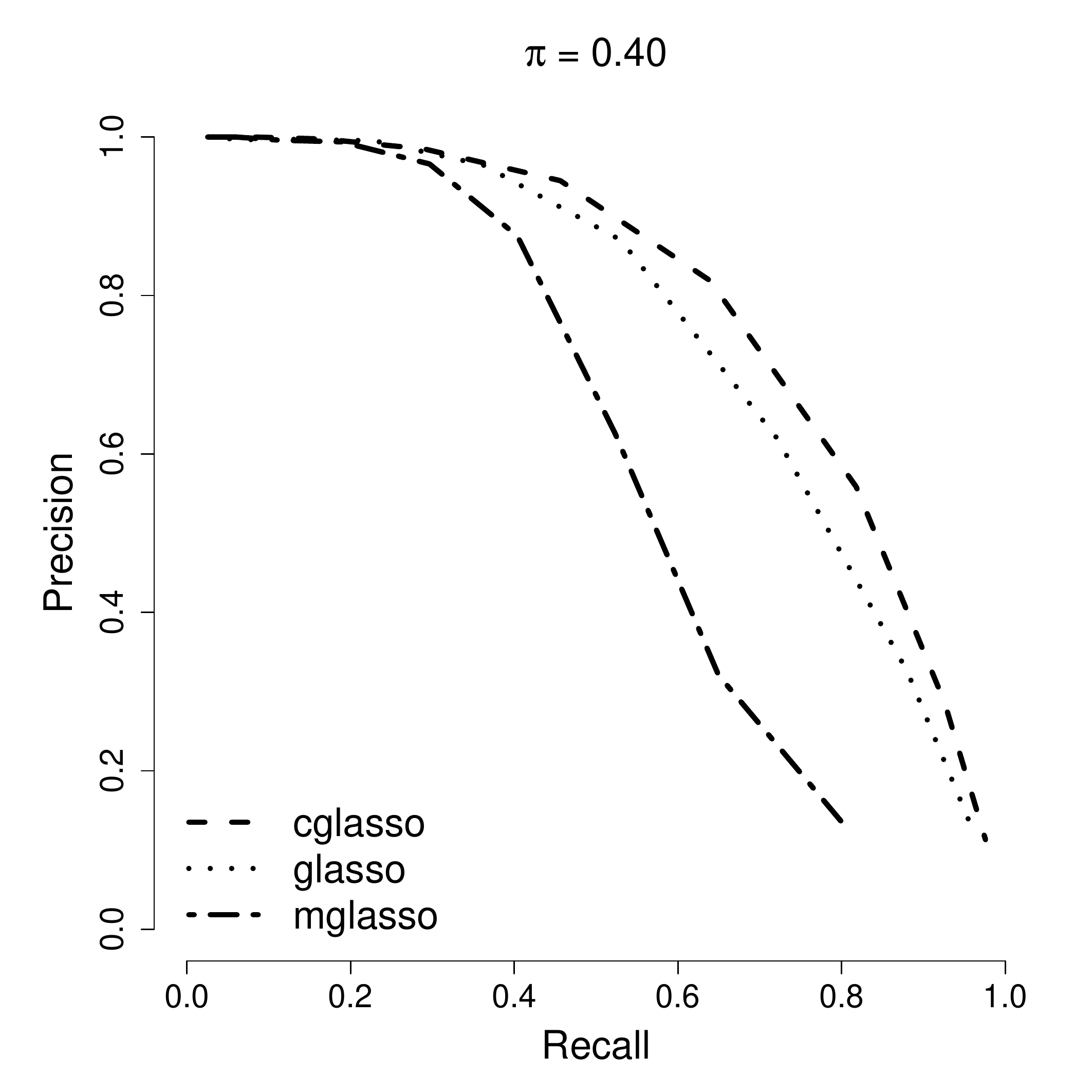}}\\
\subfigure[Precision-Recall Curves for $\widehat\Theta$ with $\lambda/\lambda_{\max} = 0.25$\label{fig:pr-curve-Tht}]
{\includegraphics[scale = 0.20]{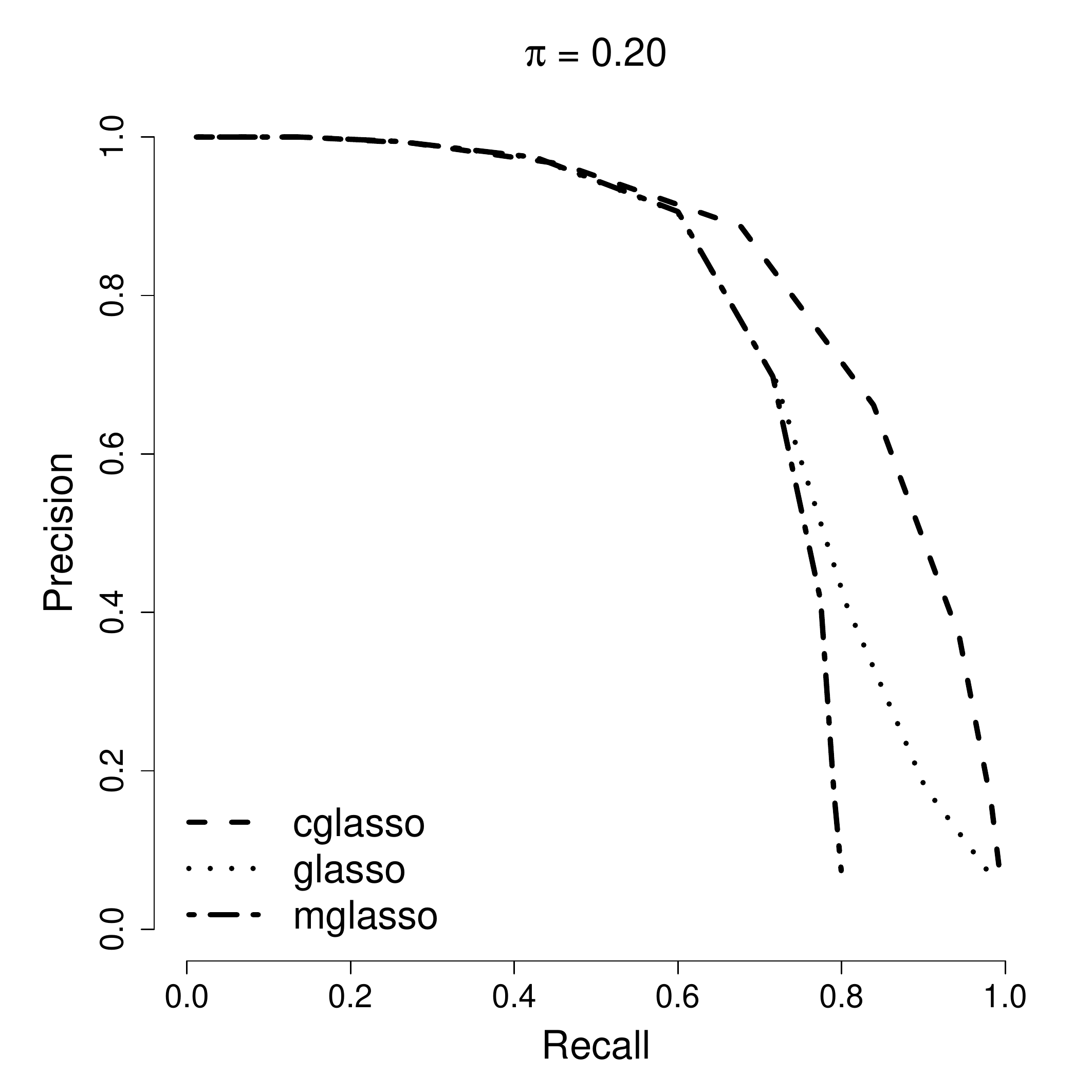}
\includegraphics[scale = 0.20]{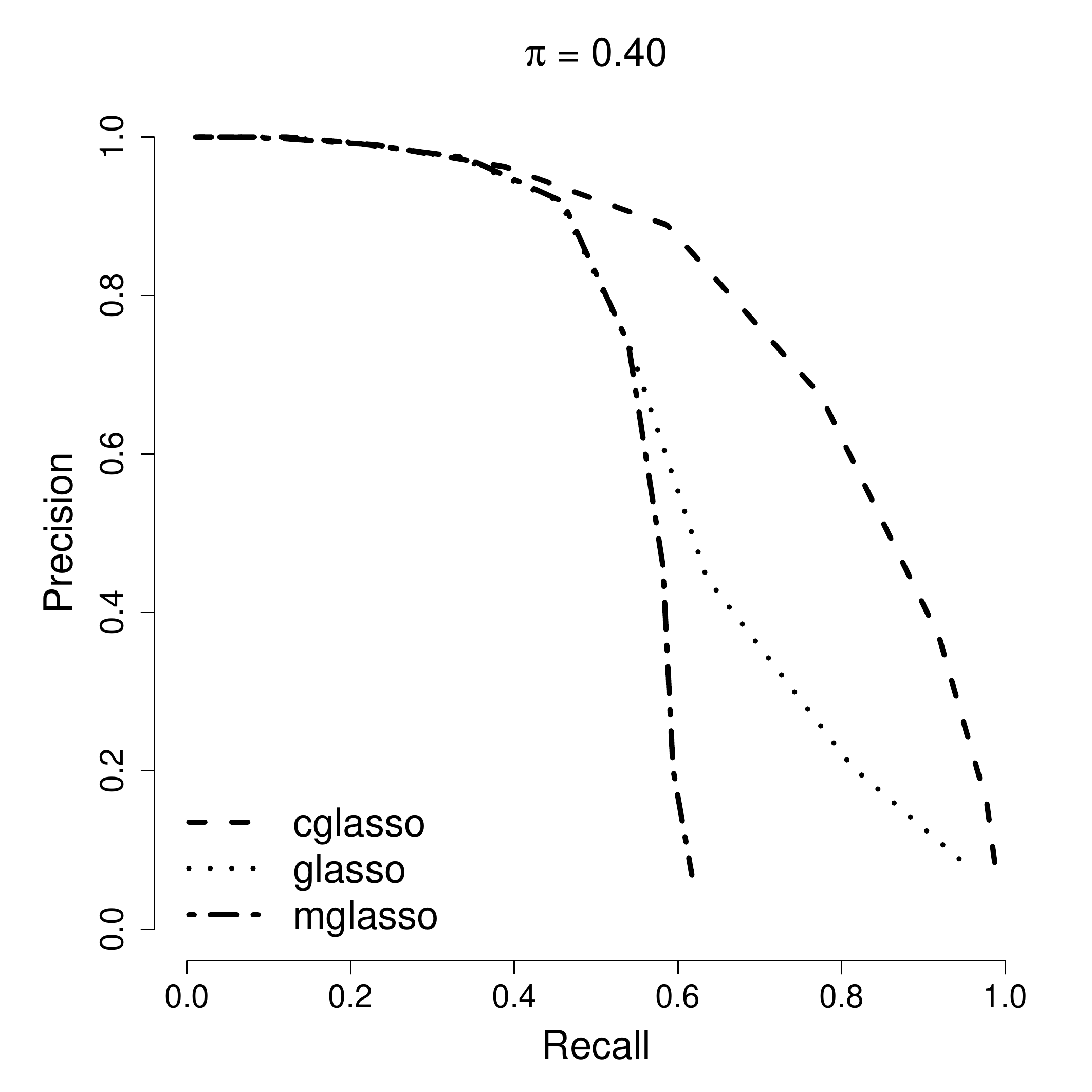}}
\caption{Precision-recall curves of the scenarios with $p$ and $q$ equal 50\label{fig:pr-curve}}
\end{figure}

Firstly, we evaluate the network recovery ability of the considered estimators. As an example, Figure~\ref{fig:pr-curve} shows the precision-recall curves of the scenarios with $p$ and $q$ equal to 50 and ratios $\rho/\rho_{\max}$ fixed to 0.25 (Figure~\ref{fig:pr-curve-B}) and $\lambda/\lambda_{\max}$ fixed to 0.25 (Figure~\ref{fig:pr-curve-Tht}).
These figures unveil two important features of the proposed estimator. Firstly, the conditional cglasso estimator behaves better  than its competitors in terms of network recovery, since, for each level of recall, it has a higher level of precision. This aspect is more pronounced in the estimation of $\Theta$. Secondly, whereas the proposed estimator is rather robust to the number of censored response variables, the results of the two competitors  depend strongly on the probability of censoring. In particular, conditional glasso and mglasso respond to an increase in the number of censored values with a reduction in terms of both precision and recall. Also in this case, the effect seems to be more pronounced for the estimation of $\Theta$ than that of $\bm{B}$.
\begin{figure}
\centering
\subfigure[Results for $\bm{\widehat{B}}$]{\includegraphics[scale = 0.45]{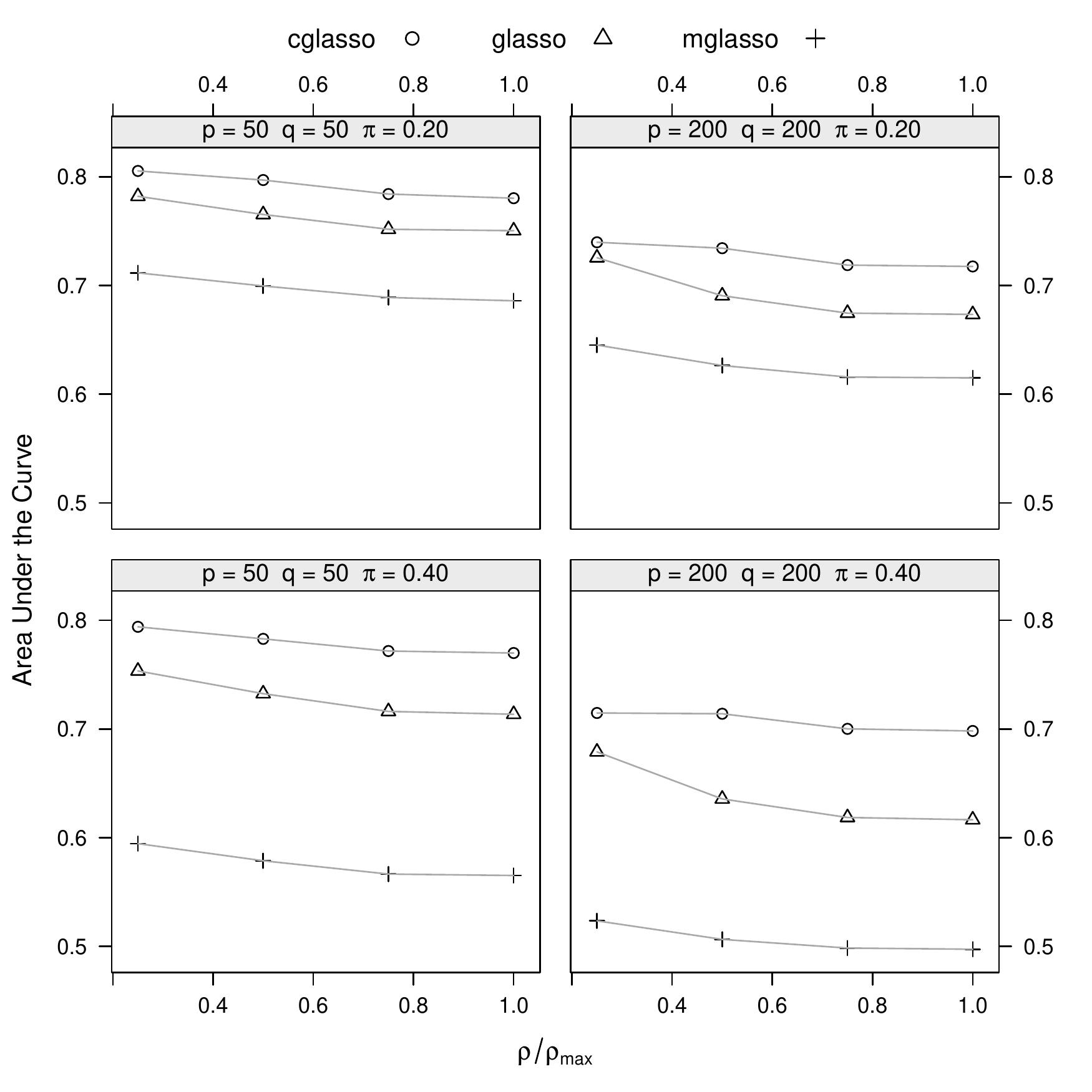}\label{fig:auc_B}}\\
\subfigure[Results for $\widehat\Theta$]{\includegraphics[scale = 0.45]{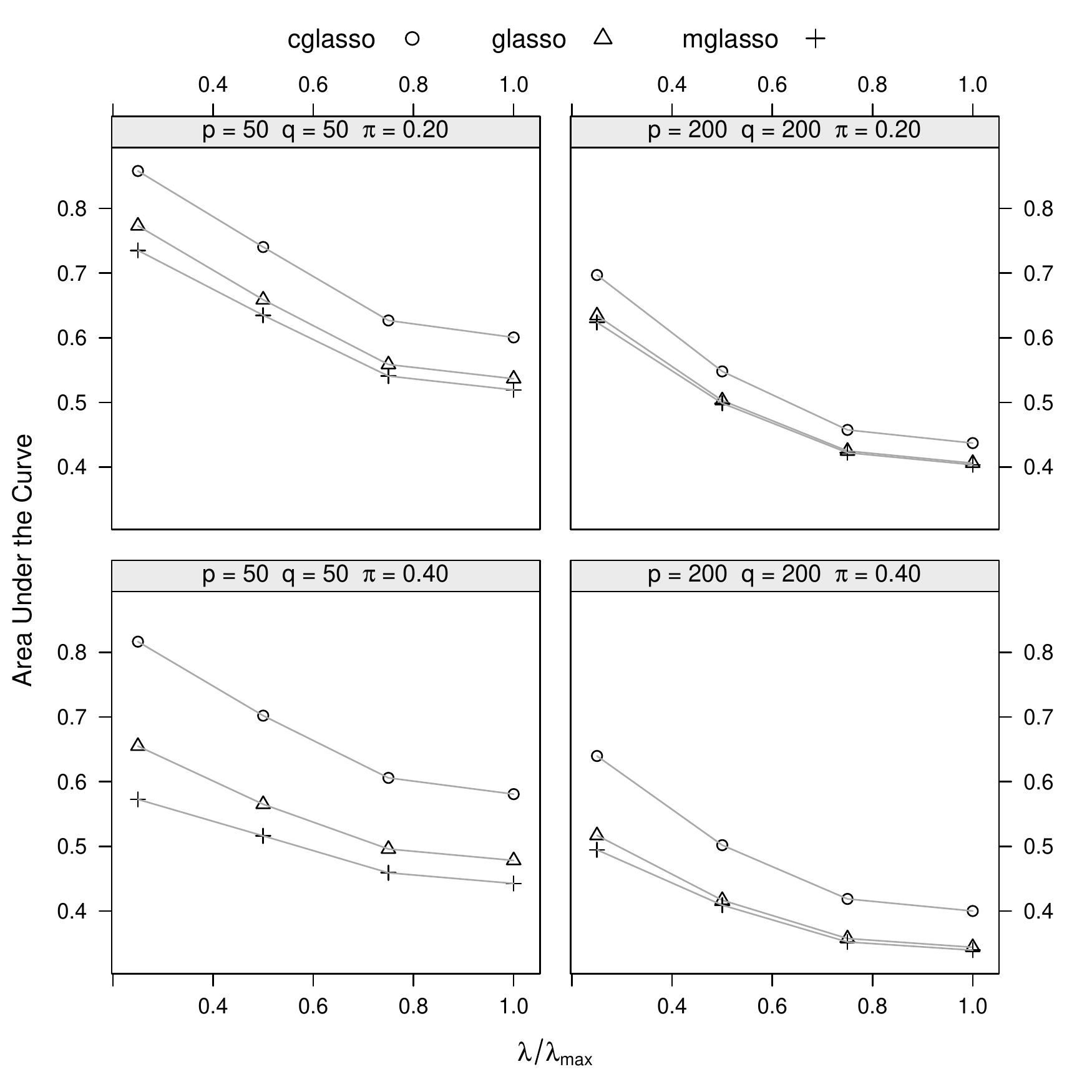}\label{fig:auc_Tht}}
\caption{Paths of the average AUCs as the ratios $\rho/\rho_{\max}$ and $\lambda/\lambda_{\max}$ are varying\label{fig:auc_path}}
\end{figure}

In order to compare the results of the low-dimensional case with the setting in which both $p$ and $q$ exceed the sample size and, at the same time, in order to study the effects of an estimator, say $\widehat\Theta$, onto the path of the other estimator, say $\bm{\widehat{B}}$, Figure~\ref{fig:auc_path} shows the average AUCs as the ratios $\rho/\rho_{\max}$ and $\lambda/\lambda_{\max}$ are varying. More specifically, Figure~\ref{fig:auc_B} shows the average AUC of the precision-recall curve of $\bm{\widehat{B}}$ as  the amount of sparsity in $\widehat\Theta$ is reduced, that is, as the ratio $\rho/\rho_{\max}$ goes from 1 to 0. In the same way, Figure~\ref{fig:auc_Tht} shows the average AUC for $\widehat\Theta$ as the shrinkage on $\bm{\widehat{B}}$ is reduced, which is controlled through $\lambda/\lambda_{\max}$. Several comments arise looking at these figures. Firstly, the initial observations coming from the analysis of Figure~\ref{fig:pr-curve} are confirmed also in the other scenarios, that is, the proposed method is preferable in terms of network recovery and is more robust to an increase in the probability of censoring. As expected, an increase in the number of variables causes a uniform loss of performance across all the  methods considered. However, also in the more challenging setting, the conditional cglasso maintains its global superiority in recovering the sparsity structure of the matrices $\bm B$ and $\Theta$. More interesting is the analysis of the crossing effects between the two estimators. Figure~\ref{fig:auc_B} suggests that the network recovery property of the three estimators of $\bm B$ is only weakly depending on the level of shrinkage on $\widehat\Theta$ since all three methods show only a slight increase in the AUC as $\rho/\rho_{\max}$ is reduced. This result is not surprising and was initially observed in~\cite{RothmanEtAl_JCGS_10} and used, by the authors, to develop an approximate method to estimate the regression coefficient matrix. Such an idea was followed also by other authors, such as \cite{YinEtAl_JMA_13} and \cite{CaiEtAl_BioK_13}. Different from the previous conclusions, Figure~\ref{fig:auc_Tht} shows that the performance of the three methods in the network recovery associated to $\widehat\Theta$  strongly depends on the ratio $\lambda/\lambda_{\max}$. This can be explained by observing that a reduction in the amount of sparsity of $\bm{\widehat{B}}$ causes an improvement in the matrix  $\bm{\widehat{S}}_{y\mid x}(\bm{\widehat{B}})$ which is translated in a better estimate of the concentration matrix and therefore an increase in the ability of network recovery.

\begin{figure}
\centering
\subfigure[MSE for $\bm{\widehat{B}}$ with $\rho/\rho_{\max} = 0.25$\label{fig_mse-curve-B}]
{\includegraphics[scale = 0.15] {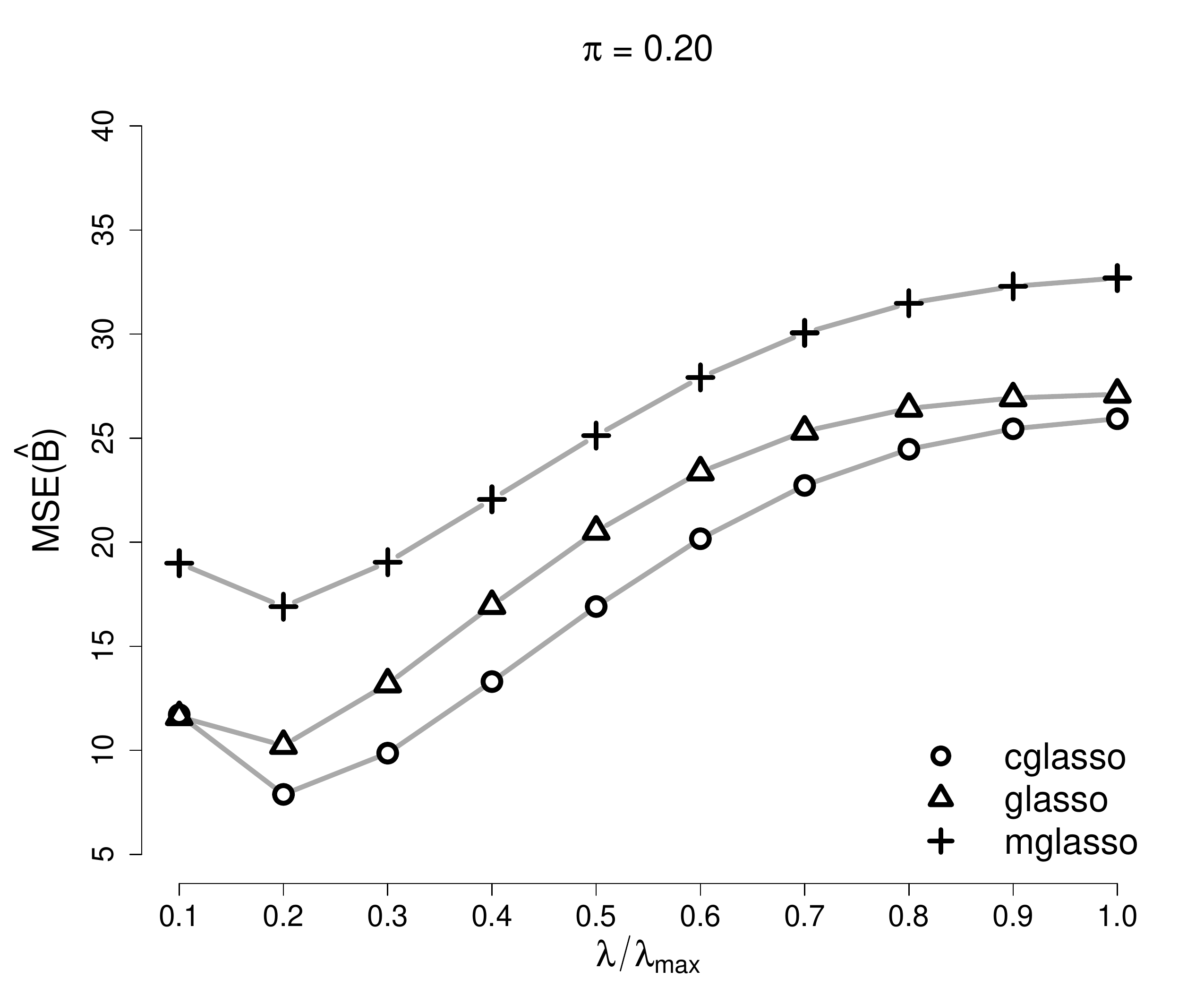} %
\includegraphics[scale = 0.15]{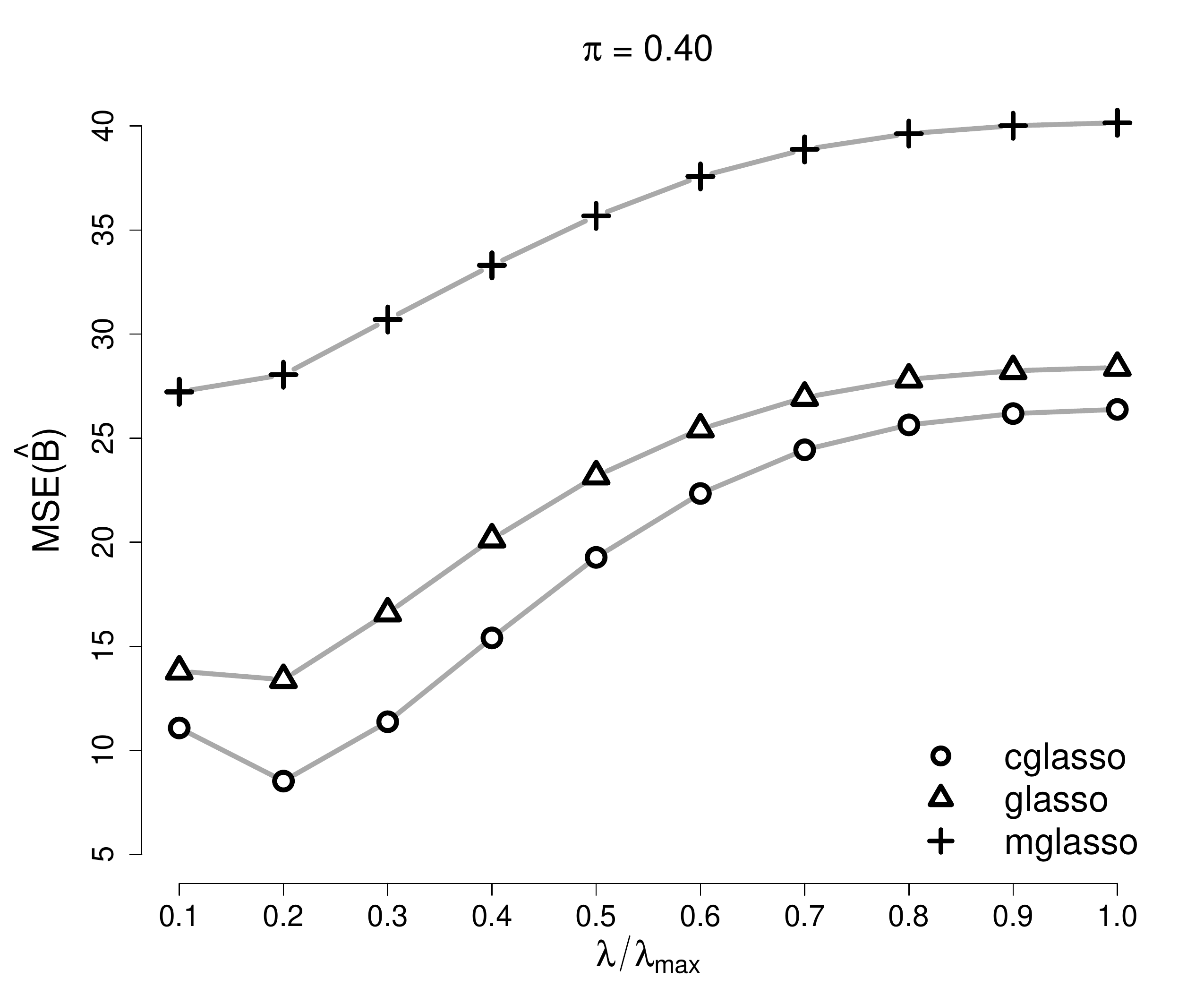}}\\
\subfigure[MSE for $\widehat\Theta$ with $\lambda/\lambda_{\max} = 0.25$\label{fig_mse-curve-Tht}]
{\includegraphics[scale = 0.15]{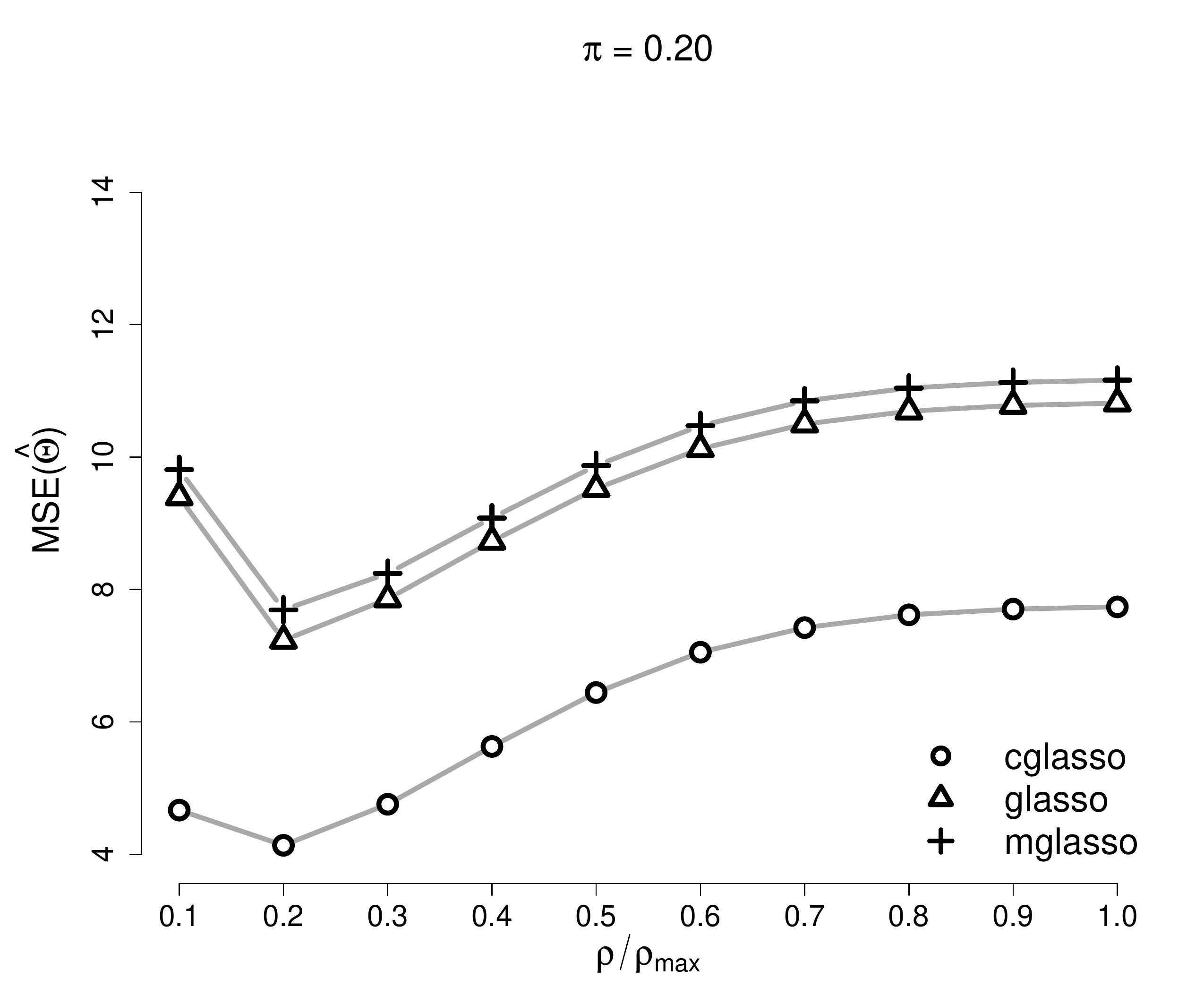} %
\includegraphics[scale = 0.15]{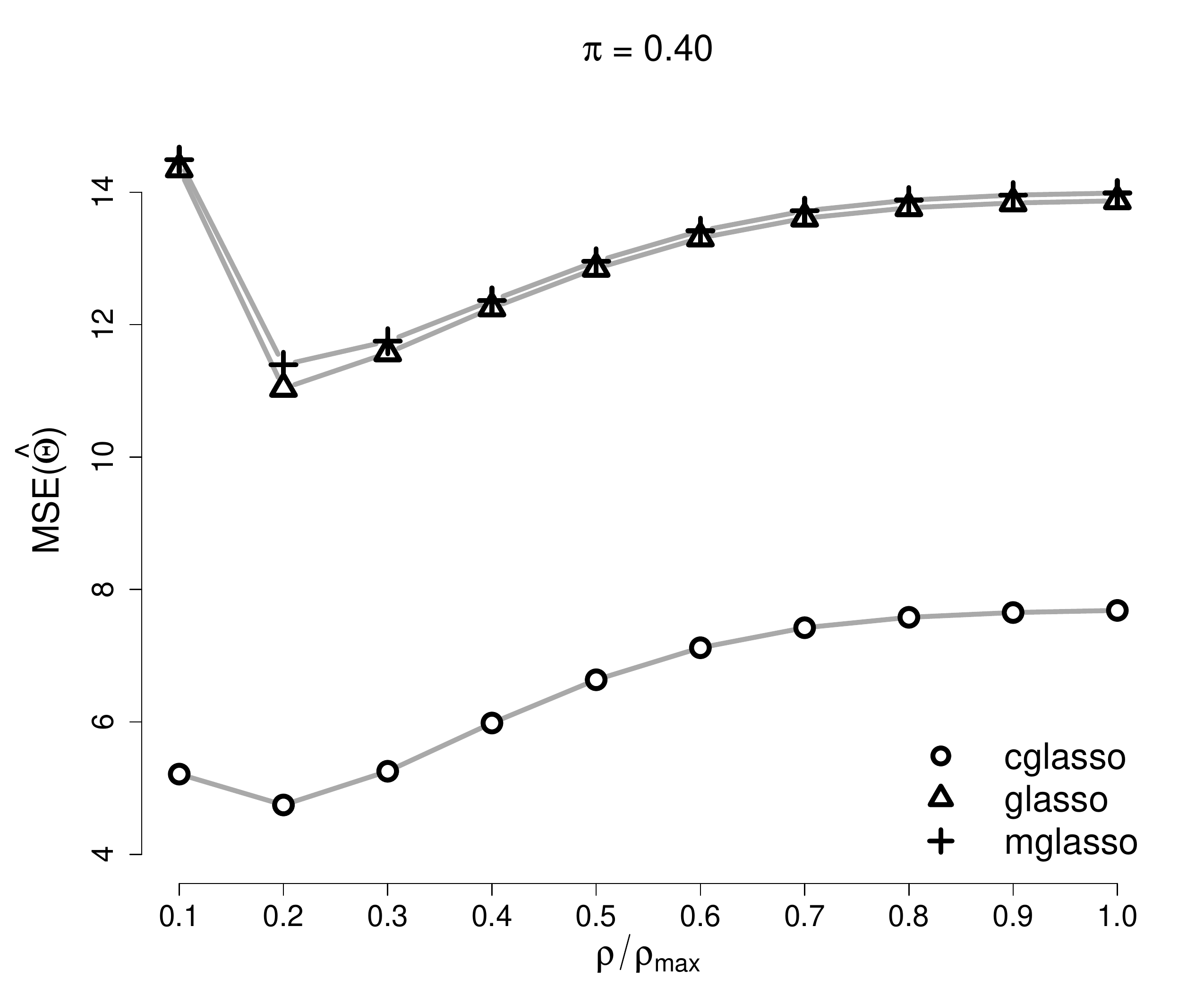}}
\caption{Paths of the mean square error for the scenario with $p$ and $q$ equal to 50\label{fig:mse-curve}}
\end{figure}

\begin{figure}
\centering
\subfigure[Results for $\bm{\widehat{B}}$]{\includegraphics[scale = 0.45]{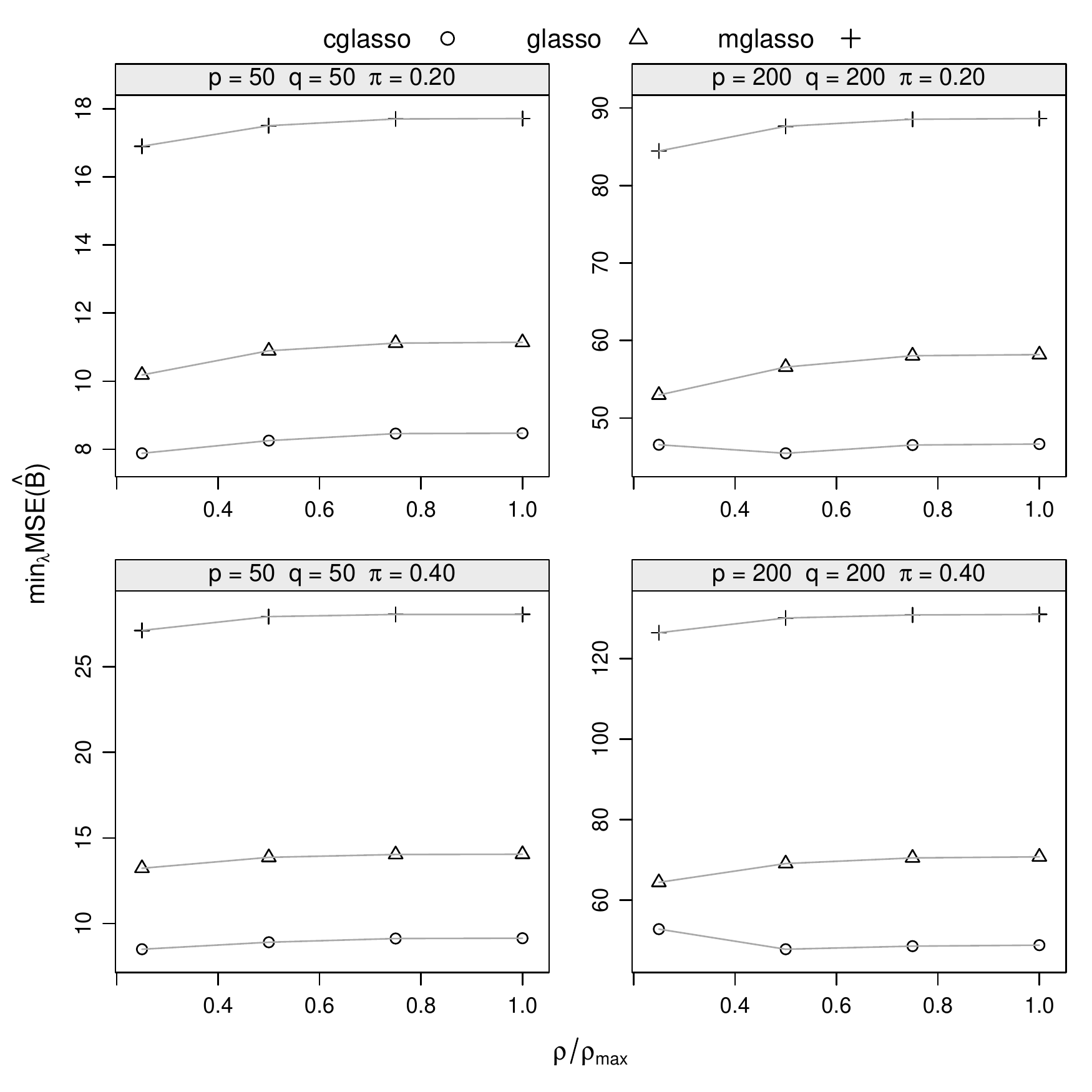}\label{fig:mse_B}}\\
\subfigure[Results for $\widehat\Theta$]{\includegraphics[scale = 0.45]{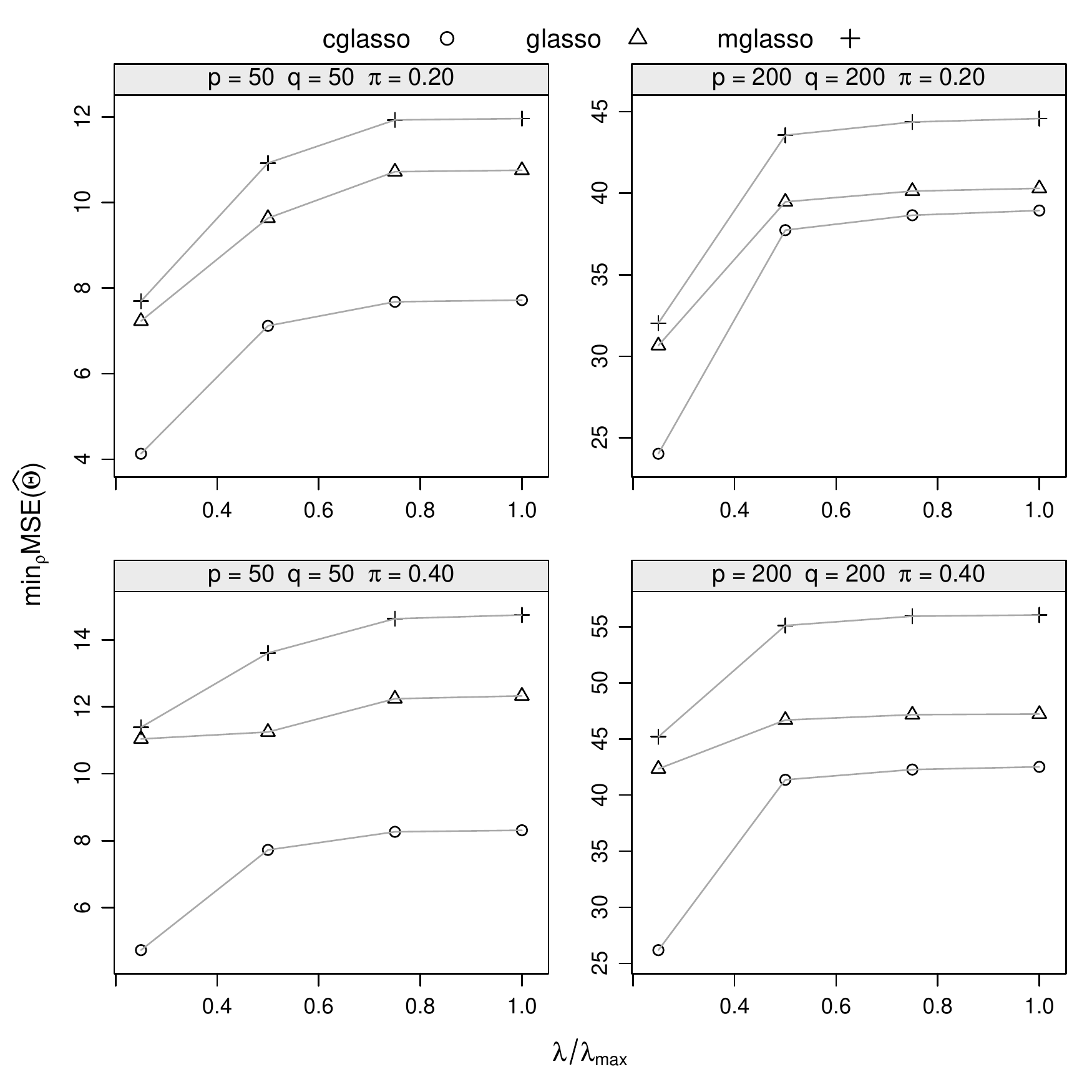}\label{fig:mse_Tht}}
\caption{Paths of the MSEs as the ratios $\rho/\rho_{\max}$ and $\lambda/\lambda_{\max}$ are varying\label{fig:mse_path}}
\end{figure}

Finally, we compare the three methods in terms of mean squared error. A quick look at Figures~\ref{fig:mse-curve} and \ref{fig:mse_path} suggests that the main conclusions, previously obtained for the network recovery case, are valid also when we study mean squared errors. Figure~\ref{fig:mse-curve} shows the average MSEs for the scenario with $p$ and $q$ equal to 50 and ratios $\rho/\rho_{\max}$ and $\lambda/\lambda_{\max}$ both fixed to 0.25. We can clearly see that the proposed conditional cglasso is globally preferred and the gap among the three methods is amplified as the probability of right censoring increases. Furthermore, also in terms of MSE, the proposed method is more robust to the percentage of censored values. In Figure~\ref{fig:mse_path} we use the minimum value of the mean squared error attained along the path to study the crossing effects among the estimators. Again, the proposed method shows a better performance, despite a global increase in MSE across all methods when we consider the high-dimensional setting.

Although our conclusions come from the analysis of the two extreme settings for $p$ and $q$, the full tables reported in the Supplementary Materials show that our method is preferable also in the intermediate settings, that is, when only one of $p$ and $q$ exceeds the sample size. This is true both in terms of AUC and MSE.

\section{The role of miRNA in multiple myeloma\label{sec:realdata}}

MicroRNAs (miRNAs) are endogenous small non-coding RNAs, approximately 22 nucleotides in length, that play a regulatory role in gene expression by mediating mRNA cleavage or translation expression~\citep{CilekEtAl_PLOSone_17}. Several studies have shown that a deregulation of the miRNAs can cause a disruption in the gene regulation mechanisms of the cell and that this might even lead to cancerous phenotypes. Furthermore, recent studies have also shown that miRNA networks have synergistic roles in the regulation of pathological conditions. If two miRNAs interact with each other in a network, they are more likely to regulate pathways and target genes which have similar functions in the same disease~\citep{CilekEtAl_PLOSone_17}.

Among these studies, \cite{GutierrezEtAl_Leukemia_10} investigated the expression level of a set of miRNAs, measured by RT-qPCR technology on a sample of patients with multiple myeloma. Multiple Myeloma (MM) is a malignant plasma cell disorder that accounts for approximately 10\% of all hematologic cancers~\citep{KyleEtAl_Blood_08}. In the study by \cite{GutierrezEtAl_Leukemia_10}, patients were selected to represent the most relevant recurrent genetic abnormalities in MM. In particular, Table~\ref{tbl:patients} reports the distribution of the cytogenetic abnormalities in the $n = 64$ experimental cases available.
\begin{table}
\begin{center}
\caption{Cytogenetic characteristics of the 64 patients for which miRNA expression data is generated.}\label{tbl:patients}
\begin{tabular}{cc}
\hline\noalign{\smallskip}
Cytogenetic Abnormalities & No. of Cases\\
\noalign{\smallskip}\hline\noalign{\smallskip}
Normal Plasma & 5\\
t(4;14) & 17\\
t(11; 14) & 11\\
t(14;16) & 4\\
RB Deletion & 14\\
Normal FISH & 13\\
\noalign{\smallskip}\hline
\multicolumn{2}{l}{Abbreviations: FISH, fluorescence in situ hybridization}
\end{tabular}
\end{center}
\end{table}
For each sample, the dataset contains the measured cycle-threshold for $p = 141$ miRNAs. As discussed in the Introduction, RT-qPCR data are typically right-censored and, in this study, the upper limit of detection was fixed to 50 cycles.

We start our analysis  by looking at the relationship between the proportion of censored data and the observed mean cycle-threshold per protein.
\begin{figure}
\centering
\includegraphics[scale = 0.36]{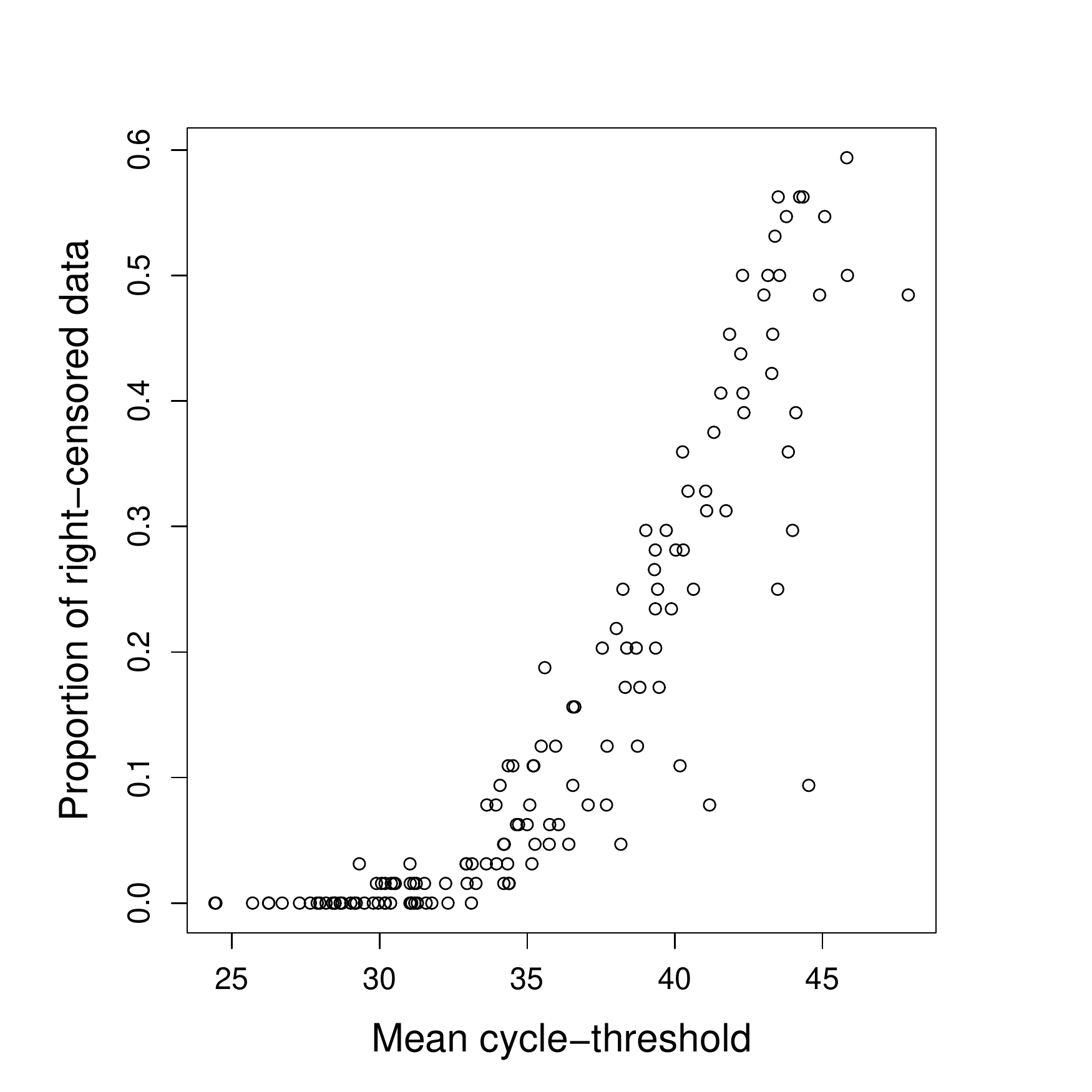}
\caption{Proportion of right-censored data versus the mean of cycle-threshold (upper limit of detection is fixed to 50 cycles).\label{fig_mm_propVSmean}}
\end{figure}
Figure~\ref{fig_mm_propVSmean} shows that the proportion of censored data is an increasing function of the mean cycle-threshold, suggesting that the assumption of missing-at-random is not suitable for this dataset. Furthermore, some miRNAs have a very high proportion of censoring, clearly highlighting the need to take the censoring mechanism into account in order to conduct a valid statistical analysis.

Using the approach proposed in this paper, we extend the study conducted in~\cite{GutierrezEtAl_Leukemia_10}, by integrating different steps of the analysis into one. In particular, we propose a model that performs differential expression, data normalization and network inference in a single step, so that, on the one hand, modelling uncertainty  can be fully accounted for, and, on the other hand, dependencies between miRNAs that can be explained by shared factors can be identified. To this end, let $\bm y_i$ be the vector of cycle-thresholds measured on the $i$-th case, $\bm z_i$ the expression level of 15 endogenous internal reference genes, called housekeeping and typically used in the normalization of RT-qPCR data~\citep{PipelersEtAl_Plos1_17,CuiEtAl_BMCGenomics_15}, and $\bm f_i$ the factor encoding the genetic abnormalities. By introducing a latent variable $\bm y^{\star}_i$, with  $\bm y_i= \mbox{min}(\bm y^{\star}_i, 50)$, we consider the model:
\begin{eqnarray*}\label{eqn:model_Ct}
E(\bm y^{\star}_i\mid \bm x_i) &=& \bm \beta_0 + \bm\beta^\top_1\bm z_i + \bm\beta^\top_2\bm f_i, \\ \nonumber
E(\bm y^{\star}_i\mid \bm x_i) &=& \Sigma_{y^{\star}_i\mid \bm x_i}
\end{eqnarray*}
where $\bm x_i = (\bm z_i^\top, \bm f_i^\top)^\top$ is a vector of $q = 20$ predictors.
 This model  takes into account simultaneously of the normalization step, via the term $\bm\beta^\top_1\bm z_i $, the effects coming from the different cytogenetic features, via the term $\bm\beta^\top_2\bm f_i$, as well as the dependencies between miRNAs, via the covariance matrix $\Sigma_{y^{\star}_i\mid \bm x_i}$ whose inverse gives the conditional independence network.

\begin{figure}
\centering
\subfigure[]{\includegraphics[scale = 0.36]{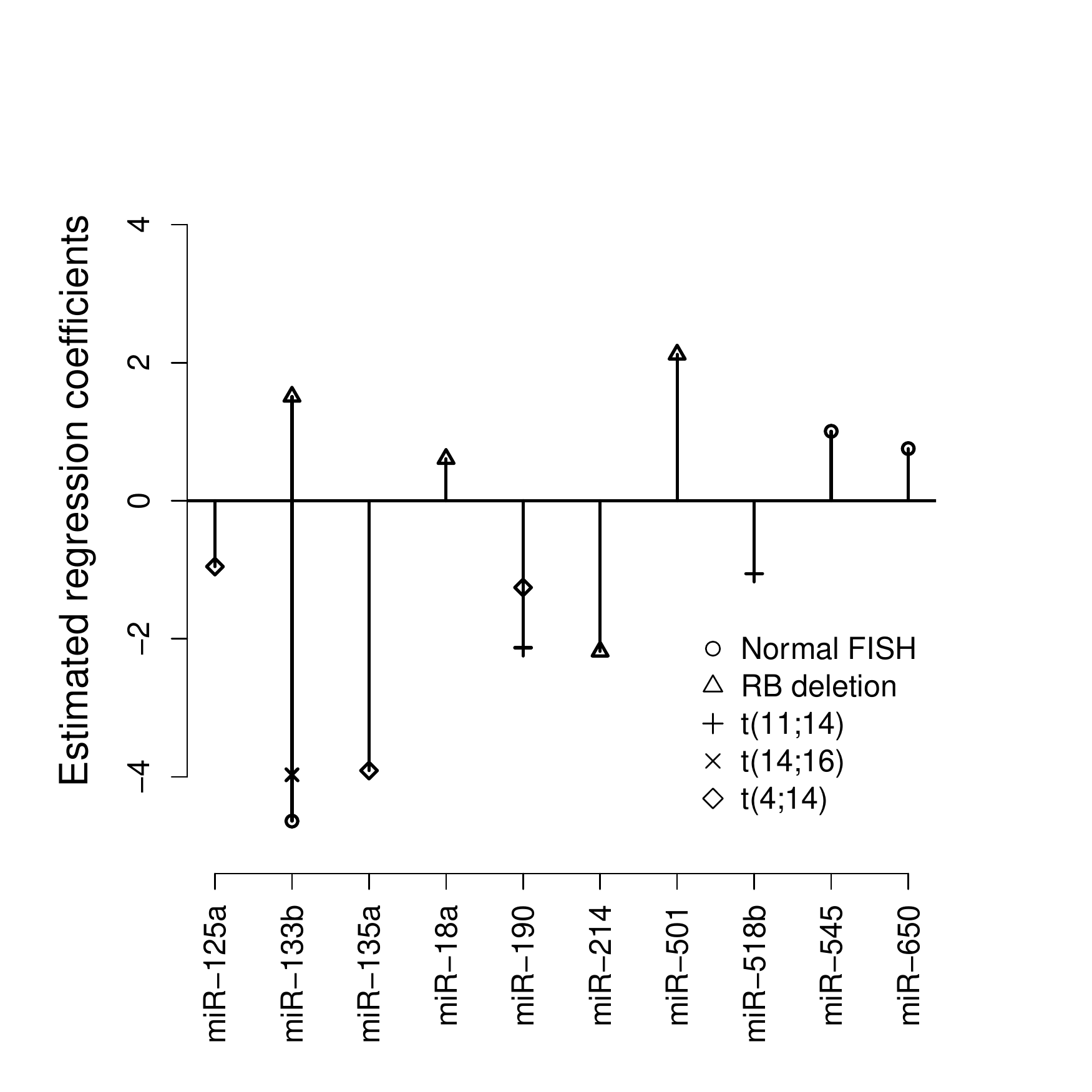}\label{fig_mm_B}}\\
\subfigure[]{\includegraphics[scale = 0.36]{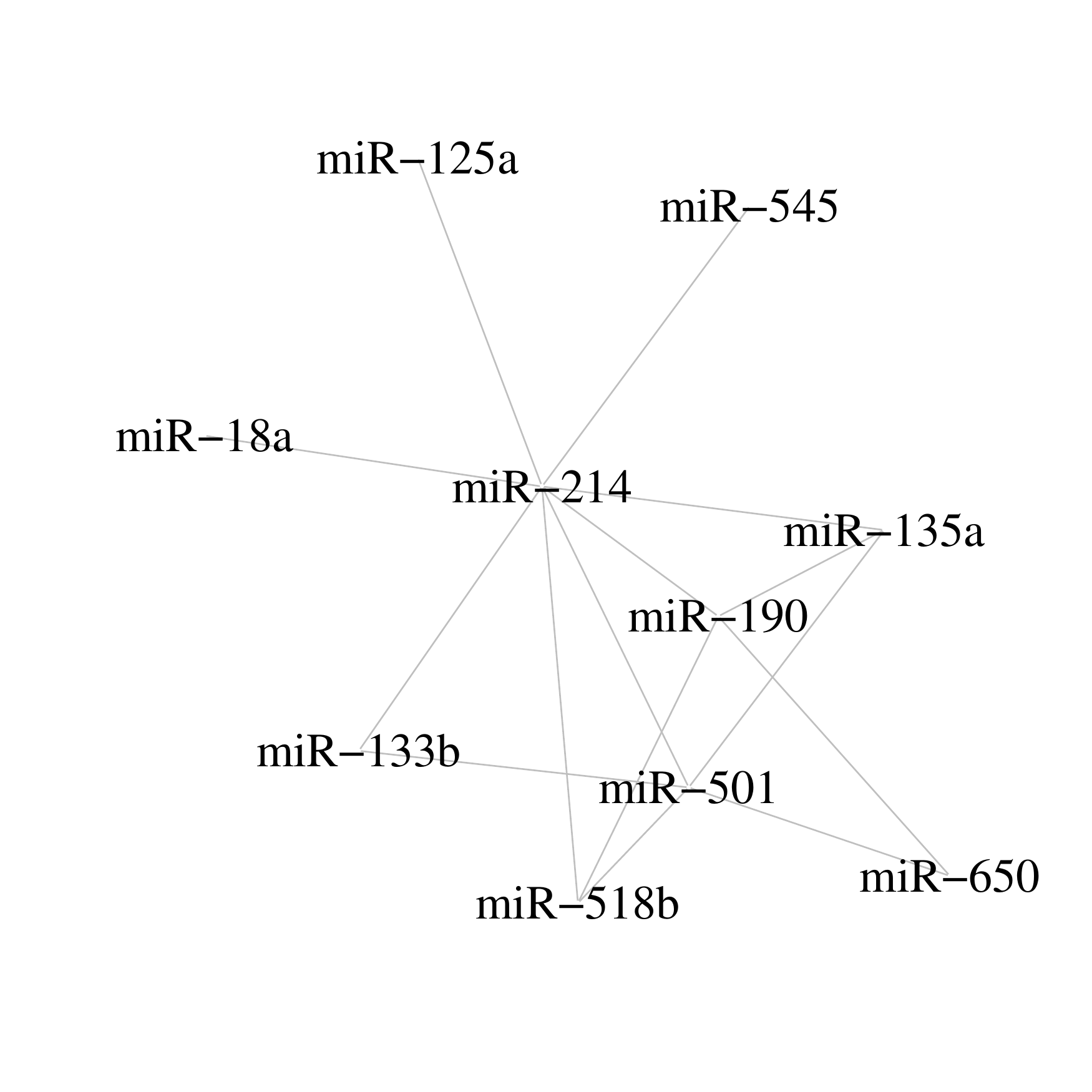}\label{fig_miRnet}}
\caption{Multiple Myeloma dataset:
(a) estimated regression coefficients of the differentially expressed miRNAs; (b): network of the differential expressed miRNAs.}
\end{figure}

 Following the procedure described in Section~\ref{sec:tuning_selection}, we select the optimal values of the two tuning parameters and we identify 90 non-zero regression coefficients and 246 non-zero partial correlation coefficients. Figure~\ref{fig_mm_B} shows the differentially expressed genes identified by the analysis when taking Normal Plasma as the reference category. Among these, miR-204 is known to have a central role in multiple myeloma pathogenesis, as reported for example by \cite{GutierrezEtAl_Leukemia_10} and \cite{Misiewicz-Krzeminska:2013aa}. On the other hand \cite{Seckinger:2015aa} elucidate the regulatory role of miR-135a in the t(4;14) translocation, whereas the role of miR-133b in the t(14;16) translocation is discussed in~\cite{Lionetti:2009aa}.
 In addition to data normalization and differential expression, which are commonly considered in this type of studies, the model proposed in this paper is also able to infer the underlying network of interactions between miRNAs, once all the other controlling factors are taken into account. Figure~\ref{fig_miRnet} shows the inferred network for the differentially expressed genes only, highlighting once more a central role for miR-204 and identifying a set of connections, which could be further investigated in terms of shared pathways and target genes.

\section{Conclusions\label{sec:conclusions}}

In this paper we have proposed a new approach to infer a sparse conditional Gaussian graphical model when data are subject to censoring. The proposed estimator, called conditional cglasso, is obtained by combining the idea of conditional glasso~\citep{YinEtAl_AOAS_11} with the censored glasso~\citep{AugugliaroEtAl_BioStat_18}. The computational problem of estimating the regression coefficient matrix and conditional independence graph of sparse conditional censored Gaussian graphical model is solved by developing a suitable EM-algorithm. The proposed algorithm is computational efficient, thanks to fast approximations of the moments of truncate Gaussian distributions at the E-step, and a two-step procedure at the M-step, which alternates graphical lasso with a novel block-coordinate descent multivariate lasso approach. The computational efficiency of the proposed approach makes it feasible for high-dimension settings. An extensive simulation study showed that the proposed estimator overcomes the existing estimators both in terms of regression coefficients estimation and of network recovery. The proposed estimator will be implemented in the \texttt{R} package \texttt{cglasso}, freely available at \url{https://CRAN.R-project.org/package=cglasso}.

\section*{Acknowledgements}
Project was partially supported by the European Cooperation in Science and Technology (COST) [COST Action CA15109 European Cooperation for Statistics of Network Data Science (COSTNET)].

\section*{Appendix}

\subsection*{Proof of Theorem~\ref{prop:main_identity}}

Before delving into technical details, we remember that the conditional distribution of $\bm y_i = (\bm y_{io_i}, \bm y_{ic_i})$ given $\bm x_i$ can be factorized as the product of two Gaussian distributions:
\begin{equation}\label{eqn:prop1}
\phi(\bm y_{io_i}, \bm y_{ic_i}\mid \bm x_i;\bm\vartheta) = \phi(\bm y_{ic_i} \mid \bm y_{io_i}, \bm x_i;\bm\vartheta_1) \phi(\bm y_{io_i} \mid, \bm x_i;\bm\vartheta_2).
\end{equation}
Let
\[
\bar\ell(\bm\vartheta) = \frac{1}{n}\sum_{i=1}^n\log \int_{D_{c_i}} \phi(\bm y_{io_i}, \bm y_{ic_i}\mid \bm x_i;\bm\vartheta) \text{d}\bm y_{ic_i},
\]
be the average observed log-likelihood function, then
\begin{align*}
\frac{\partial\bar\ell(\bm\vartheta)}{\partial\vartheta_{hk}} = & \frac{1}{n}\sum_{i = 1}^n\frac{\int_{D_{c_i}} \partial\phi(\bm y_{io_i}, \bm y_{ic_i}\mid \bm x_i;\bm\vartheta)/\partial\vartheta_{hk}\,\text{d}\bm y_{ic_i}}{\int_{D_{c_i}} \phi(\bm y_{io_i}, \bm y_{ic_i}\mid \bm x_i;\bm\vartheta)\,\text{d}\bm y_{ic_i}} \\
= & \frac{1}{n}\sum_{i = 1}^n \int_{D_{c_i}} \frac{\partial\log \phi(\bm y_{io_i}, \bm y_{ic_i}\mid \bm x_i;\bm\vartheta)}{\partial\vartheta_{hk}}\times \frac{\phi(\bm y_{io_i}, \bm y_{ic_i}\mid \bm x_i;\bm\vartheta)}{\int_{D_{c_i}} \phi(\bm y_{io_i}, \bm y_{ic_i}\mid \bm x_i;\bm\vartheta)\,\text{d}\bm y_{ic_i}}\text{d}\bm y_{ic_i}.
\end{align*}
Using factorization~(\ref{eqn:prop1}), the previous expression can be simplified in the following way:
\begin{equation*}
\frac{\partial\bar\ell(\bm\vartheta)}{\partial\vartheta_{hk}} = \frac{1}{n}\sum_{i = 1}^n \int_{D_{c_i}} \frac{\partial\log \phi(\bm y_{io_i}, \bm y_{ic_i}\mid \bm x_i;\bm\vartheta)}{\partial\vartheta_{hk}}  \times \frac{\phi(\bm y_{ic_i} \mid \bm y_{io_i}, \bm x_i;\bm\vartheta_1)}{\int_{D_{c_i}} \phi(\bm y_{ic_i} \mid \bm y_{io_i}, \bm x_i;\bm\vartheta_1)}\text{d}\bm y_{ic_i}.
\end{equation*}
Since $\phi(\bm y_{ic_i} \mid \bm y_{io_i}, \bm x_i;\bm\vartheta_1) / \int_{D_{c_i}} \phi(\bm y_{ic_i} \mid \bm y_{io_i}, \bm x_i;\bm\vartheta_1)$ is the density function of the conditional distribution of $\bm y_{ic_i}$ given $\{\bm x_i, \bm y_{io_i}\}$, which is a Gaussian distribution truncated over $D_{c_i}$,  we have that $\partial \bar\ell(\bm\vartheta)/\partial\vartheta_{hk}$ is equal to:
\[
\frac{1}{n}\sum_{i= 1}^n E\left(\frac{\partial\log\phi(\bm y_{io_i}, \bm y_{ic_i}\mid \bm x_i;\bm\vartheta)}{\partial\vartheta_{hk}} \mid \bm y_{ic_i}\in D_{c_i},\bm x_i;\bm\vartheta\right).
\]
This complete the proof.

\subsection*{Proof of Theorem~\ref{prop:multi_lasso}}

We start the proof by noting that
\begin{align}\label{eqn:prop2_1}
\mathrm{tr}\{\widehat\Theta\bm{\widehat{S}}_{y\mid x}(\bm B)\}  = & \frac{1}{n} \mathrm{tr}\{\widehat\Theta(\bm{\widehat C}_{yy} - \bm{\widehat C}_{yx} \bm B - \bm B^\top\bm{\widehat C}_{xy} + \bm B^\top \bm C_{xx} \bm B)\}\nonumber \\
= & \frac{1}{n} \mathrm{tr}\{\widehat\Theta(\widehat{\bmY}^\top\widehat{\bmY} - \bm{\widehat C}_{yx} \bm B - \bm B^\top\bm{\widehat C}_{xy} + \bm B^\top \bm C_{xx} \bm B)\} + \frac{1}{n} \mathrm{tr}\{\widehat\Theta(\bm{\widehat C}_{yy} - \widehat{\bmY}^\top\widehat{\bmY})\}\nonumber \\
 = & \frac{1}{n}\mathrm{tr}\{\widehat\Theta(\widehat{\bmY} - \bmX\bm B)^\top(\widehat{\bmY} - \bmX\bm B)\} + \frac{1}{n} \mathrm{tr}\{\widehat\Theta(\bm{\widehat C}_{yy} - \widehat{\bmY}^\top\widehat{\bmY})\}.
\end{align}
Now, suppose that we are interested in fitting the $k$th column of the regression coefficient matrix while the remaining columns are held fixed to the current estimate, then we can define the following partition
$
\bm B =
\begin{pmatrix}
\bm B_k & \bm{\widehat{B}}_{\setminus k}
\end{pmatrix},
$
where $\bm{\widehat{B}}_{\setminus k}$ denotes the matrix of the current estimates. Furthermore, let $\bmR_{\setminus k} = \widehat{\bmY}_{\setminus k} - \bmX_{\setminus k} \bm{\widehat{B}}_{\setminus k}$ denote the matrix of the current residuals. With such notation we have
\begin{equation*}
(\widehat{\bmY} - \bmX\bm B)^\top( \widehat{\bmY}  -  \bmX\bm B) =  \begin{pmatrix}
\|\widehat{\bmY}_k - \bmX\bm B_k\|^2 & (\widehat{\bmY}_k - \bmX\bm B_k)^\top\bmR_{\setminus k}\\
\bmR_{\setminus k}^\top(\widehat{\bmY}_k - \bmX\bm B_k) & \bmR_{\setminus k}^\top\bmR_{\setminus k}
\end{pmatrix},
\end{equation*}
\[
\widehat\Theta = %
\begin{pmatrix}
\hat\theta_{kk} & \widehat\Theta_{k\setminus k} \\
 \widehat\Theta_{\setminus k k} &  \widehat\Theta_{\setminus k\setminus k}
\end{pmatrix},
\]
then, using the identity~(\ref{eqn:prop2_1}), the trace term in the following minimization problem
\begin{equation}\label{eqn:prop2_2}
\min_{\bm B_k}  \mathrm{tr}\{\widehat\Theta\bm{\widehat{S}}_{y\mid x}(\bm B_k)\} + \lambda  \hat\theta_{kk} \|\bm\beta_k\|_1,
\end{equation}
can be written as follows
\begin{align*}
\mathrm{tr}\{\widehat\Theta\bm{\widehat{S}}_{y\mid x}(\bm B_k)\}   = & \frac{\hat\theta_{kk}}{n}\{\|\widehat{\bmY}_k - \bmX\bm B_k\|^2  + 2 (\widehat{\bmY}_k - \bmX\bm B_k)^\top\widehat{\bmV}_k\} + \frac{1}{n} \mathrm{tr}\{\widehat\Theta_{\setminus k\setminus k} \bmR_{\setminus k}^\top\bmR_{\setminus k}\} + \frac{1}{n} \mathrm{tr}\{\widehat\Theta(\bm{\widehat C}_{yy} - \widehat{\bmY}^\top\widehat{\bmY})\}\nonumber \\
= & \frac{\hat\theta_{kk}}{n}\|\widehat{\bmY}_k + \widehat{\bmV}_k - \bmX\bm B_k\|^2 + \mathrm{C},
\end{align*}
where
\begin{align*}
\widehat{\bmV}_k = & \hat\theta_{kk}^{-1}\bmR_{\setminus k}\widehat\Theta_{\setminus k k}\\
\mathrm{C} = & n^{-1}[\mathrm{tr}\{\widehat\Theta_{\setminus k\setminus k} \bmR_{\setminus k}^\top\bmR_{\setminus k}\} +  \mathrm{tr}\{\widehat\Theta(\bm{\widehat C}_{yy} - \widehat{\bmY}^\top\widehat{\bmY})\}  - \hat\theta_{kk} \|\widehat{\bmV}_k \|^2]
\end{align*}
are constant terms with respect to $\bm B_k$. Finally, letting $\widetilde\bmY_k = \widehat{\bmY}_k + \widehat{\bmV}_k$, it is easy to see that minimization problem~(\ref{eqn:prop2_2}) is equivalent to
\[
\min_{\bm B_k} \frac{1}{n} \|\widetilde\bmY_k - \bmX\bm B_k \|^2  + \lambda\|\bm\beta_k\|_1.
\]
This complete the proof.


\subsection*{Proof of Theorem~\ref{prop:max_lambdarho}}
The proof is divided in two steps; first, assuming known the pair $(\lambda_{\max}, \rho_{\max})$, we explain how to compute the corresponding conditional cglasso estimators then, analyzing the Karush-Kuhn-Tucker (KKT) conditions, we derive the exact fomulae to compute the maximum value of the two tuning parameters.

Suppose that $\lambda_{\max}$ and $\rho_{\max}$ are known then $\bm{\widehat B = (\bm{\hat\beta}_0, \bm{\hat\beta}^\top})^\top$ and $\widehat\Theta$ are maximally penalized, that is $\bm{\hat\beta} = \bm 0$ and $\hat\theta_{hk} = 0$ for any $h\ne k$. In this case, since the remaining parameters are unpenalized they can be estimated by simply maximizing the following average observed log-likelihood function:
\begin{equation*}
\frac{1}{n}\sum_{k = 1}^p   \left\{ \sum_{i\in o_k} \log\phi(y_{ik};\mu_k,\sigma^2_k) +  \right. + |c^{-}_k|\log\int_{-\infty}^{l_k} \phi(y;\mu_k,\sigma^2_k) \text{d} y  + \left. |c^{+}_k|\log\int_{u_k}^{+\infty} \phi(y;\mu_k,\sigma^2_k) \text{d} y\right\},
\end{equation*}
where $\bm{\hat\beta}_0 =  (\hat\mu_1,\ldots,\hat\mu_k)^\top$ and $\widehat\Theta = \diag{\hat\sigma^{-2}_1,\ldots,\hat\sigma^{-2}_p}$. Since the previous average log-likelihood function is defined through the sum of $p$ marginal log-likelihood functions, the pair $\{\hat\mu_k, \hat\sigma^2_k\}$ is merely  the solution of the following maximization problem:
\begin{equation*}
\max_{\mu_k,\sigma^2_k} \sum_{i\in o_k} \log\phi(y_{ik};\mu_k,\sigma^2_k)  + |c^{-}_k|\log\int_{-\infty}^{l_k} \phi(y;\mu_k,\sigma^2_k) \text{d} y  + |c^{+}_k|\log\int_{u_k}^{+\infty} \phi(y;\mu_k,\sigma^2_k) \text{d} y.
\end{equation*}
In order to compute $\lambda_{\max}$ and $\rho_{\max}$, first we derive the KKT conditions characterizing the solution of the following maximization problem
\[
\max \bar\ell(\bm B, \Theta) - \lambda\sum_{k = 1}^p\theta_{kk}\|\bm\beta_k\|_1 - \rho\|\Theta\|_1^-.
\]
In what follows, we shall use the convention that a column of a matrix is denoted indexing the matrix with the corresponding index; for example, $\bmX_k$ denotes the $k$th column of the matrix $\bmX$. Furthermore, the generic entry of the matrix
$\widehat S_{y\mid x}(\bm{\widehat{B}})$ is denoted by $\hat s_{hk}(\bm{\widehat B})$.

By straightforward algebra and using the results given in Theorem~\ref{prop:main_identity}, for a given pair $(\lambda, \rho)$ the KKT conditions can be written in matrix form as:
\begin{align*}
 \frac{1}{n} \sum_{l = 1}^p\hat\theta_{kl} \bmX_h^\top(\widehat\bmY_l - \bmX\bm{\widehat B}_l)- \lambda\hat\theta_{kk} \nu(\hat\beta_{hk}) &= 0,\\
\hat\sigma_{hk}^2 - \hat s_{hk}(\bm{\widehat B}) - \rho\nu(\hat\theta_{hk}) &= 0,
\end{align*}
where, in general, $\nu(\hat\vartheta_{hk})$ denotes the subgradient of the absolute value function at $\hat\vartheta_{hk}$, that is, $\nu(\hat\vartheta_{hk}) = \sign{\hat\vartheta_{hk}}$ if and only if $\hat\vartheta_{hk}\ne0$ otherwise $|\nu(\hat\vartheta_{hk})|\le1$. Remembering that at $(\lambda_{\max}, \rho_{\max})$ we that $\bm\beta_0 = \bm{\hat\mu}$, $\bm{\hat\beta} = \bm 0$ and $\widehat\Theta = \diag{\hat\sigma^{-2}_1,\ldots,\hat\sigma^{-2}_p}$, the previous conditions can be simplified as follows:
\begin{align*}
\frac{1}{n} \sum_{i = 1}^m x_{ih}(\hat y_{i,k} - \hat\mu_k) &= \lambda_{\max} \nu(\hat\beta_{hk}),\\
\hat s_{hk}(\bm{\widehat B}) &= - \rho_{\max}\nu(\hat\theta_{hk}).
\end{align*}
Finally, taking the absolute value of both sides of the previous equations and remembering that $|\nu(\hat\vartheta_{hk})|\le1$ we have
\begin{align*}
\frac{1}{n} |\sum_{i = 1}^m x_{ih}(\hat y_{i,k} - \hat\mu_k)| = \lambda_{\max} |\nu(\hat\beta_{hk})|\le\lambda_{\max},\\
|\hat s_{hk}(\bm{\widehat B})| = \rho_{\max}|\nu(\hat\theta_{hk})|\le\rho_{\max},
\end{align*}
from which it follows immediately that
\begin{align*}
\lambda_{\max} = & \max_{h,k} n^{-1} |\sum_{i = 1}^m x_{ih}(\hat y_{i,k} - \hat\mu_k)| \\
\rho_{\max} = & \|\widehat S_{y\mid x}(\bm{\widehat{B}})\|_\infty^-,
\end{align*}
where $\|\widehat S_{y\mid x}(\bm{\widehat{B}})\|_\infty^- = \max_{h\ne k} |\hat s_{hk}(\bm{\widehat B})|$.


\bibliographystyle{chicago}
\bibliography{conditional-cglassobib}

\begin{thebibliography}{}

\bibitem[\protect\citeauthoryear{Augugliaro, Abbruzzo, and
  Vinciotti}{Augugliaro et~al.}{2018}]{AugugliaroEtAl_BioStat_18}
Augugliaro, L., A.~Abbruzzo, and V.~Vinciotti (2018).
\newblock The $\ell_1$-penalized censored {Gaussian} graphical model.
\newblock {\em Biostatistics\/}.

\bibitem[\protect\citeauthoryear{Cai, Li, Liu, and Xie}{Cai
  et~al.}{2013}]{CaiEtAl_BioK_13}
Cai, T., H.~Li, W.~Liu, and J.~Xie (2013).
\newblock Covariate-adjusted precision matrix estimation with an application in
  genetical genomics.
\newblock {\em Biometrika\/}~{\em 100}, 139--156.

\bibitem[\protect\citeauthoryear{Cai and Liu}{Cai and
  Liu}{2011}]{CaiEtAl_JASA_11}
Cai, T. and W.~Liu (2011).
\newblock Adaptive thresholding for sparse covariance matrix estimation.
\newblock {\em Journal of the American Statistical Association\/}~{\em
  106\/}(494), 672--684.

\bibitem[\protect\citeauthoryear{Chen, Ren, Zhao, and Zhou}{Chen
  et~al.}{2016}]{ChenEtAl_JASA_16}
Chen, M., Z.~Ren, H.~Zhao, and H.~Zhou (2016).
\newblock Asymptotically normal and efficient estimation of covariate-adjusted
  {Gaussian} graphical model.
\newblock {\em Journal of the American Statistical Association\/}~{\em
  111\/}(513), 394--406.

\bibitem[\protect\citeauthoryear{Chiquet, Huard, and Robin}{Chiquet
  et~al.}{2017}]{ChiquetEtAl_StatComp_17}
Chiquet, J., T.~M. Huard, and S.~Robin (2017).
\newblock Structured regularization for conditional {Gaussian} graphical
  models.
\newblock {\em Statistics and Computing\/}~{\em 27\/}(3), 789--805.

\bibitem[\protect\citeauthoryear{Cilek, Ozturk, and Dedeoglu}{Cilek
  et~al.}{2017}]{CilekEtAl_PLOSone_17}
Cilek, E.~E., H.~Ozturk, and B.~G. Dedeoglu (2017).
\newblock Construction of {miRNA}-{miRNA} networks revealing the complexity of
  { miRNA}-mediated mechanisms in trastuzumab treated breast cancer cell lines.
\newblock {\em PLoS One\/}~{\em 12\/}(10), e0185558.

\bibitem[\protect\citeauthoryear{Cui, Yu, Tamhane, Causey, Steg, Danila,
  Reynolds, Wang, Wanzeck, Tang, Ledbetter, Redden, Johnson, and Jr}{Cui
  et~al.}{2015}]{CuiEtAl_BMCGenomics_15}
Cui, X., S.~Yu, A.~Tamhane, Z.~L. Causey, A.~Steg, M.~I. Danila, R.~J.
  Reynolds, J.~Wang, K.~C. Wanzeck, Q.~Tang, S.~S. Ledbetter, D.~T. Redden,
  M.~R. Johnson, and L.~B. Jr (2015).
\newblock Simple regression for correcting $\delta c_t$ bias in {RT-qPCR}
  low-density array data normalization.
\newblock {\em BMC Bioinformatics\/}~{\em 16\/}(1), 82--93.

\bibitem[\protect\citeauthoryear{Dempster, Laird, and Rubin}{Dempster
  et~al.}{1977}]{DempsterEtAl_JRSSB_77}
Dempster, A.~P., N.~M. Laird, and D.~B. Rubin (1977).
\newblock Maximum likelihood from incomplete data via the {EM} algorithm.
\newblock {\em Journal of the Royal Statistical Society. Series B\/}~{\em
  39\/}(1), 1--38.

\bibitem[\protect\citeauthoryear{Derveaux, Vandesompele, and
  Hellemans}{Derveaux et~al.}{2010}]{DerveauxEtAl_Methods_10}
Derveaux, S., J.~Vandesompele, and J.~Hellemans (2010).
\newblock How to do successful gene expression analysis using real-time {PCR}.
\newblock {\em Methods\/}~{\em 50\/}(4), 227--230.

\bibitem[\protect\citeauthoryear{Friedman, Hastie, and Tibshirani}{Friedman
  et~al.}{2008}]{FriedmanEtAl_biostat_08}
Friedman, J.~H., T.~Hastie, and R.~Tibshirani (2008).
\newblock Sparse inverse covariance estimation with the graphical lasso.
\newblock {\em Biostatistics\/}~{\em 9\/}(3), 432--441.

\bibitem[\protect\citeauthoryear{Friedman, Hastie, and Tibshirani}{Friedman
  et~al.}{2010}]{FriedmanEtAl_JSS_10}
Friedman, J.~H., T.~Hastie, and R.~Tibshirani (2010).
\newblock Regularization paths for generalized linear models via coordinate
  descent.
\newblock {\em Journal of Statistical Software\/}~{\em 33\/}(1), 1--22.

\bibitem[\protect\citeauthoryear{Friedman, Hastie, and Tibshirani}{Friedman
  et~al.}{2018}]{glasso_manual}
Friedman, J.~H., T.~Hastie, and R.~Tibshirani (2018).
\newblock {\em glasso: Graphical lasso- estimation of {Gaussian} graphical
  models}.

\bibitem[\protect\citeauthoryear{Guo, Levina, Michailidis, and Zhu}{Guo
  et~al.}{2015}]{GuoEtAl_JCGS_15}
Guo, J., E.~Levina, G.~Michailidis, and J.~Zhu (2015).
\newblock Graphical models for ordinal data.
\newblock {\em Journal of Computational and Graphical Statistics\/}~{\em
  24\/}(1), 183--204.

\bibitem[\protect\citeauthoryear{Guti{\'e}rrez, Sarasquete,
  Misiewicz-Krzeminska, Delgado, Rivas, Ticona, Ferm{i\~n}\'{a}n,
  Mart{\'\i}n-Jim{\'e}nez, Chill{\'o}n, Risue{\~n}o, Hern{\'a}ndez,
  Garc{\'\i}a-Sanz, Gonz{\'a}lez, and Miguel}{Guti{\'e}rrez
  et~al.}{2010}]{GutierrezEtAl_Leukemia_10}
Guti{\'e}rrez, N., M.~Sarasquete, I.~Misiewicz-Krzeminska, M.~Delgado, J.~D.~L.
  Rivas, F.~Ticona, E.~Ferm{i\~n}\'{a}n, P.~Mart{\'\i}n-Jim{\'e}nez,
  C.~Chill{\'o}n, A.~Risue{\~n}o, J.~Hern{\'a}ndez, R.~Garc{\'\i}a-Sanz,
  M.~Gonz{\'a}lez, and J.~S. Miguel (2010).
\newblock Deregulation of {microRNA} expression in the different genetic
  subtypes of multiple myeloma and correlation with gene expression profiling.
\newblock {\em Leukemia\/}~{\em 24\/}(3), 629--637.

\bibitem[\protect\citeauthoryear{Huang and Chen}{Huang and
  Chen}{2018}]{HuangEtAl_IEEE_TKDE_18}
Huang, F. and S.~Chen (2018).
\newblock Learning dynamic conditional {Gaussian} graphical models.
\newblock {\em IEEE Transactions on Knowledge and Data Engineering\/}~{\em
  30\/}(4), 703--716.

\bibitem[\protect\citeauthoryear{Huang, Songcan, and Huang}{Huang
  et~al.}{2018}]{HuangEtAl_IEEE_TNNLS_18}
Huang, F., Songcan, and S.-J. Huang (2018).
\newblock Joint estimation of multiple conditional {Gaussian} graphical models.
\newblock {\em IEEE Transactions on Neural Networks and Learning
  Systems\/}~{\em 29\/}(7), 3034--3046.

\bibitem[\protect\citeauthoryear{Ibrahim, Zhu, and Tang}{Ibrahim
  et~al.}{2008}]{IbrahimEtAl_JASA_08}
Ibrahim, J.~G., H.~Zhu, and N.~Tang (2008).
\newblock Model selection criteria for missing-data problems using the {EM}
  algorithm.
\newblock {\em Journal of the American Statistical Association\/}~{\em
  103\/}(484), 1648--1658.

\bibitem[\protect\citeauthoryear{Kyle and Rajkumar}{Kyle and
  Rajkumar}{2008}]{KyleEtAl_Blood_08}
Kyle, R.~A. and V.~S. Rajkumar (2008).
\newblock Multiple myeloma.
\newblock {\em Blood\/}~{\em 111\/}(6), 2962--2972.

\bibitem[\protect\citeauthoryear{Lafferty, McCallum, and Pereira}{Lafferty
  et~al.}{2001}]{LaffertyEtAl_ICML_01}
Lafferty, J., A.~McCallum, and F.~C. Pereira (2001).
\newblock Conditional random fields: Probabilistic models for segmenting and
  labeling sequence data.
\newblock In {\em Proceedings of the 18th International Conference on Machine
  Learning 2001 (ICML 2001)}, pp.\  282--289.

\bibitem[\protect\citeauthoryear{Lauritzen}{Lauritzen}{1996}]{Lauritzen_book_96}
Lauritzen, S.~L. (1996).
\newblock {\em Graphical Models}.
\newblock Oxford: Oxford University Press.

\bibitem[\protect\citeauthoryear{Lee and Liu}{Lee and
  Liu}{2012}]{LeeEtAl_JMA_2012}
Lee, W. and Y.~Liu (2012).
\newblock Simultaneous multiple response regression and inverse covariance
  matrix estimation via penalized {Gaussian} maximum likelihood.
\newblock {\em J. Multivariate Anal.\/}~{\em 11}, 241--255.

\bibitem[\protect\citeauthoryear{Li, Chun, and Zhao}{Li
  et~al.}{2012}]{LiEtAl_JASA_12}
Li, B., H.~Chun, and H.~Zhao (2012).
\newblock Sparse estimation of conditional graphical models with application to
  gene networks.
\newblock {\em Journal of the American Statistical Association\/}~{\em
  107\/}(497), 152--167.

\bibitem[\protect\citeauthoryear{Lionetti, Biasiolo, Agnelli, Todoerti, Mosca,
  Fabris, Sales, Deliliers, Bicciato, Lombardi, Bortoluzzi, and Neri}{Lionetti
  et~al.}{2009}]{Lionetti:2009aa}
Lionetti, M., M.~Biasiolo, L.~Agnelli, K.~Todoerti, L.~Mosca, S.~Fabris,
  G.~Sales, G.~Deliliers, S.~Bicciato, L.~Lombardi, S.~Bortoluzzi, and A.~Neri
  (2009).
\newblock Identification of {microRNA} expression patterns and definition of a
  {microRNA/mRNA} regulatory network in distinct molecular groups of multiple
  myeloma.
\newblock {\em Blood\/}~{\em 114\/}(25), e20--6.

\bibitem[\protect\citeauthoryear{Little and Rubin}{Little and
  Rubin}{2002}]{LittleEtAl_Book_02}
Little, R. J.~A. and D.~B. Rubin (2002).
\newblock {\em Statistical Analysis with Missing Data\/} (second edition ed.).
\newblock Hoboken, NJ, USA: John Wiley \& Sons, Inc.

\bibitem[\protect\citeauthoryear{McCall, McMurray, Land, and Almudevar}{McCall
  et~al.}{2014}]{McCallEtAl_BioInfo_14}
McCall, M.~N., H.~R. McMurray, H.~Land, and A.~Almudevar (2014).
\newblock On non-detects in {qPCR} data.
\newblock {\em Bioinformatics\/}~{\em 30\/}(16), 2310--2316.

\bibitem[\protect\citeauthoryear{McLachlan and Krishnan}{McLachlan and
  Krishnan}{2008}]{McLachlanEtAl_Book_08}
McLachlan, G. and T.~Krishnan (2008).
\newblock {\em The {EM} Algorithm and Extensions\/} (second edition ed.).
\newblock Hoboken, NJ, USA: John Wiley \& Sons, Inc.

\bibitem[\protect\citeauthoryear{{Misiewicz-Krzeminska}, {Sarasquete},
  Quwaider, Krzeminski, Ticona, Pa{\'\i}no, Delgado, Aires, Ocio,
  {Garc{\'\i}a-Sanz}, {San Miguel}, and Guti{\'e}rrez}{{Misiewicz-Krzeminska}
  et~al.}{2013}]{Misiewicz-Krzeminska:2013aa}
{Misiewicz-Krzeminska}, I., M.~E. {Sarasquete}, D.~Quwaider, P.~Krzeminski,
  F.~Ticona, T.~Pa{\'\i}no, M.~Delgado, A.~Aires, E.~M. Ocio,
  R.~{Garc{\'\i}a-Sanz}, J.~F. {San Miguel}, and N.~C. Guti{\'e}rrez (2013).
\newblock Restoration of {microRNA-214} expression reduces growth of myeloma
  cells through positive regulation of {P53} and inhibition of {DNA}
  replication.
\newblock {\em Haematologica\/}~{\em 98\/}(4), 640--648.

\bibitem[\protect\citeauthoryear{Pipelers, Clement, Vynck, Hellemans,
  Vandesompele, and Thas}{Pipelers et~al.}{2017}]{PipelersEtAl_Plos1_17}
Pipelers, P., L.~Clement, M.~Vynck, J.~Hellemans, J.~Vandesompele, and O.~Thas
  (2017).
\newblock A unified censored normal regression model for {qPCR} differential
  gene expression analysis.
\newblock {\em PLoS One\/}~{\em 12\/}(8), e0182832.

\bibitem[\protect\citeauthoryear{Rothman, Levina, and Zhu}{Rothman
  et~al.}{2010}]{RothmanEtAl_JCGS_10}
Rothman, A.~J., E.~Levina, and J.~Zhu (2010).
\newblock Sparse multivariate regression with covariance estimation.
\newblock {\em Journal of Computational and Graphical Statistics\/}~{\em
  19\/}(4), 947--962.

\bibitem[\protect\citeauthoryear{Seckinger, Mei{\ss}ner, Moreaux, Benes,
  Hillengass, Castoldi, Zimmermann, Ho, Jauch, Goldschmidt, Klein, and
  Hose}{Seckinger et~al.}{2015}]{Seckinger:2015aa}
Seckinger, A., T.~Mei{\ss}ner, T.~Moreaux, V.~Benes, J.~Hillengass,
  M.~Castoldi, J.~Zimmermann, A.~Ho, A.~Jauch, H.~Goldschmidt, B.~Klein, and
  D.~Hose (2015).
\newblock {miRNAs} in multiple myeloma--a survival relevant complex regulator
  of gene expression.
\newblock {\em Oncotarget\/}~{\em 6\/}(36), 39165--39183.

\bibitem[\protect\citeauthoryear{Sohn and Kim}{Sohn and
  Kim}{2012}]{SohnEtAl_PICAIS_12}
Sohn, K.-A. and S.~Kim (2012, 21--23 Apr).
\newblock Joint estimation of structured sparsity and output structure in
  multiple-output regression via inverse-covariance regularization.
\newblock In N.~D. Lawrence and M.~Girolami (Eds.), {\em Proceedings of the
  Fifteenth International Conference on Artificial Intelligence and
  Statistics}, Volume~22 of {\em Proceedings of Machine Learning Research}, La
  Palma, Canary Islands, pp.\  1081--1089. PMLR.

\bibitem[\protect\citeauthoryear{St\"{a}dler and B\"{u}hlmann}{St\"{a}dler and
  B\"{u}hlmann}{2012}]{StadlerEtAl_StatComp_12}
St\"{a}dler, N. and P.~B\"{u}hlmann (2012).
\newblock Missing values: sparse inverse covariance estimation and an extension
  to sparse regression.
\newblock {\em Statistics and Computing\/}~{\em 22\/}(1), 219--235.

\bibitem[\protect\citeauthoryear{Tibshirani}{Tibshirani}{1996}]{Tibshirani_JRSSB_96}
Tibshirani, R. (1996).
\newblock Regression shrinkage and selection via the lasso.
\newblock {\em Journal of the Royal Statistical Society. Series B\/}~{\em
  58\/}(1), 267--288.

\bibitem[\protect\citeauthoryear{Tseng}{Tseng}{2001}]{Tseng_JOTA_01}
Tseng, P. (2001).
\newblock Convergence of a block coordinate descent method for
  nondifferentiable minimization.
\newblock {\em J. Optim. Theory Appl.\/}~{\em 109\/}(3), 475--494.

\bibitem[\protect\citeauthoryear{Wang}{Wang}{2015}]{Wang_StatSinica_15}
Wang, J. (2015).
\newblock Joint estimation of sparse multivariate regression and conditional
  graphical models.
\newblock {\em Statist. Sinica\/}~{\em 25}, 831--851.

\bibitem[\protect\citeauthoryear{Witten, Friedman, and Simon}{Witten
  et~al.}{2011}]{WittenEtAl_JCGS_11}
Witten, D.~M., J.~H. Friedman, and N.~Simon (2011).
\newblock New insights and faster computations for the graphical lasso.
\newblock {\em Journal of Computational and Graphical Statistics\/}~{\em
  20\/}(4), 892--900.

\bibitem[\protect\citeauthoryear{Wytock and Kolter}{Wytock and
  Kolter}{2013}]{WytockEtAl_pmlr_13}
Wytock, M. and Z.~Kolter (2013).
\newblock Sparse {Gaussian} conditional random fields: Algorithms, theory, and
  application to energy forecasting.
\newblock In S.~Dasgupta and D.~McAllester (Eds.), {\em Proceedings of the 30th
  International Conference on Machine Learning}, Volume~28, pp.\  1265--1273.

\bibitem[\protect\citeauthoryear{Yin and Li}{Yin and
  Li}{2011}]{YinEtAl_AOAS_11}
Yin, J. and H.~Li (2011).
\newblock A sparse conditional {Gaussian} graphical model for analysis of
  genetical genomics data.
\newblock {\em The Annals of Applied Statistics\/}~{\em 5\/}(4), 2630--2650.

\bibitem[\protect\citeauthoryear{Yin and Li}{Yin and Li}{2013}]{YinEtAl_JMA_13}
Yin, J. and H.~Li (2013).
\newblock Adjusting for high-dimensional covariates in sparse precision matrix
  estimation by $\ell_1$-penalization.
\newblock {\em J. Multivariate Anal.\/}~{\em 116}, 365--381.

\bibitem[\protect\citeauthoryear{Yuan and Lin}{Yuan and
  Lin}{2007}]{YuanEtAl_BioK_07}
Yuan, M. and Y.~Lin (2007).
\newblock Model selection and estimation in the {Gaussian} graphical model.
\newblock {\em Biometrika\/}~{\em 94\/}(1), 19--35.

\bibitem[\protect\citeauthoryear{Yuan and Zhang}{Yuan and
  Zhang}{2014}]{YuanEtAl_IEEETIT_14}
Yuan, X.~T. and T.~Zhang (2014).
\newblock Partial {Gaussian} graphical model estimation.
\newblock {\em IEEE Transactions on Information Theory\/}~{\em 60\/}(3),
  1673--1687.

\bibitem[\protect\citeauthoryear{Zhang and Kim}{Zhang and
  Kim}{2014}]{ZhangEtAl_PLOS_14}
Zhang, L. and S.~Kim (2014).
\newblock Learning gene networks under {SNP} perturbations using {eQTL}
  datasets.
\newblock {\em PLOS Computational Biology\/}~{\em 10\/}(2), 1--20.

\bibitem[\protect\citeauthoryear{Zhao, Li, Liu, Roeder, Lafferty, and
  Wasserman}{Zhao et~al.}{2015}]{huge_manual}
Zhao, T., X.~Li, H.~Liu, K.~Roeder, J.~Lafferty, and L.~Wasserman (2015).
\newblock {\em huge: High-Dimensional Undirected Graph Estimation}.
\newblock \texttt{R} package version 1.2.7.

\end{thebibliography}

\end{document}